\documentclass[a4paper,fleqn,usenatbib]{mnras}

\usepackage{newtxtext,newtxmath}

\usepackage[T1]{fontenc}
\usepackage{ae,aecompl}
\usepackage{xcolor}

\usepackage{graphicx}	
\usepackage{amsmath}	
\usepackage[section]{placeins}
\usepackage{subfig}
\usepackage{float}

\newcommand{\hMpc}{{\ifmmode{~h^{-1}{\rm Mpc}}\else{$h^{-1}$Mpc}\fi}}
\newcommand{\hkpc}{{\ifmmode{~h^{-1}{\rm kpc}}\else{$h^{-1}$kpc}\fi}}
\newcommand{\hGpc}{{\ifmmode{~h^{-1}{\rm Gpc}}\else{$h^{-1}$Gpc}\fi}}
\newcommand{\hMsun}{{\ifmmode{h^{-1}{\rm {M_{\odot}}}}\else{$h^{-1}{\rm{M_{\odot}}}$}\fi}}
\newcommand{\Msun}{{\ifmmode{{\rm {M_{\odot}}}}\else{${\rm{M_{\odot}}}$}\fi}}

\newcommand{\halos}{haloes}

\newcommand{\rockstar}{\textsc{Rockstar}}
\newcommand{\ahf}{\textsc{AHF}}
\newcommand{\disperse}{\textsc{DisPerse}}

\newcommand{\gadgetx}{\textsc{Gadget-X}}
\newcommand{\threehundred}{\textsc{The ThreeHundred }}

\title[Filaments in the outskirts of galaxy clusters]{\textsc{The ThreeHundred}: the structure and properties of cosmic filaments in the outskirts of galaxy clusters}

\author[Rost et al.]{\parbox{\textwidth}{
Agustin Rost,$^{1,2}$\thanks{E-mail: (AR)} 
Ulrike Kuchner,$^{2}$ Charlotte Welker,$^{10,4}$ Frazer Pearce,$^{2}$\\
Federico Stasyszyn,$^{1}$ Meghan Gray,$^{2}$ Weiguang Cui,$^{5}$\\ 
Romeel Dave,$^{5,6}$ Alexander Knebe,$^{3,7,8}$ Gustavo Yepes,$^{7,8}$ Elena Rasia $^{9}$\\
}
\vspace{0.4cm}
\\
\parbox{\textwidth}{
$^1$Instituto de Astronom\'ia Te\'orica y Experimental (IATE), Laprida 854, C\'ordoba, Argentina\\
$^2$School of Physics \& Astronomy, University of Nottingham, Nottingham NG7 2RD, UK\\
$^3$International Centre for Radio Astronomy Research, The University of Western Australia, 35 Stirling Highway, Crawley, Western Australia 6009, Australia\\
$^4$McMaster University, Canada\\
$^5$Institute for Astronomy, University of Edinburgh, Royal Observatory, Edinburgh EH9 3HJ, United Kingdom\\
$^6$University of the Western Cape, Bellville, Cape Town 7535, South Africa\\
$^7$Departamento de F\'isica Te\'{o}rica, M\'{o}dulo 15, Facultad de Ciencias, Universidad Aut\'{o}noma de Madrid, 28049 Madrid, Spain\\
$^8$Centro de Investigaci\'{o}n Avanzada en F\'{\i}sica Fundamental (CIAFF), Universidad Aut\'{o}noma de Madrid, 28049 Madrid, Spain \\
$^9$ INAF, Osservatorio Astronomico di Trieste, via Tiepolo 11, I-34131, Trieste, Italy \\
$^{10}$The Johns Hopkins University, Krieger School of Arts and Science, Department of physics and Astronomy, Baltimore, MD, USA\\ 
}}
\date{Accepted XXX. Received YYY; in original form ZZZ}

\pubyear{2015}

\begin{document}
\label{firstpage}
\pagerange{\pageref{firstpage}--\pageref{lastpage}}
\maketitle

\begin{abstract}
Galaxy cluster outskirts are described by complex velocity fields induced by diffuse material collapsing towards filaments, gas and galaxies falling into clusters, and gas shock processes triggered by substructures.
A simple scenario that describes the large-scale tidal fields of the cosmic web is not able to fully account for this variety, nor for the differences between gas and collisionless dark matter.
We have studied the filamentary structure in zoom-in resimulations centred on 324 clusters from \threehundred project, focusing on differences between dark and baryonic matter.
This paper describes the properties of filaments around clusters out to five $R_{200}$, based on the diffuse filament medium where haloes had been removed. For this, we stack the remaining particles of all simulated volumes to calculate the average profiles of dark matter and gas filaments.
We find that filaments increase their thickness closer to nodes and detect signatures of gas turbulence at a distance of $\sim 2 \rm{\hMpc}$ from the cluster. These are absent in dark matter. 
Both gas and dark matter collapse towards filament spines at a rate of $\sim 200 \rm{km ~ s^{-1} h^{-1}} $. We see that gas preferentially enters the cluster as part of filaments, and leaves the cluster centre outside filaments. We further see evidence for an accretion shock just outside the cluster. For dark matter, this preference is less obvious. We argue that this difference is related to the turbulent environment. This indicates that filaments act as highways to fuel the inner regions of clusters with gas and galaxies.
\end{abstract}



\section{Introduction}
\label{intro}
Matter in the Universe is structured as a complex network called the cosmic web \citep{Bond1996}. In this picture, galaxy clusters mark the high density nodes of the network, connected by a multitude of elongated filaments and walls that, together, host most of the mass in the Universe. In contrast, cosmic voids occupy the majority of the volume, but are relatively depleted of material \citep{Forero_Romero2009, Cautun, Tempel}.
Cosmological simulations of the formation of this large-scale structure suggest a sequence in which dark matter and gas are gravitationally attracted from under-dense regions (3-dimensional voids) to increasingly denser regions of 2-dimensional walls, which contract into 1-dimensional filaments through which the material flows actracted by the large potential well of (clusters located in) the nodes.
The underlying physics of this structure formation is based on theoretical work by \citet{Zeldovich1970} who recognized the role of the large-scale tidal field as the major driving force in shaping the cosmic web, based on the initial density perturbations in the early Universe 
\citep{Arnold1982, Shandarin1984, Gurbatov1989, Shandarin1989, Hidding2014}. 

Today, approximately $40\%$ of the baryons in the Universe are expected to consist of gas at temperatures of $10^{5} - 10^{7}\rm{K}$, the warm hot intergalactic medium (WHIM), which is investigated as a possible solution of the so-called {\it missing baryon problem} \citep{Ostriker, Dave, Reimers2002, Bykov2008}.  
To a large extent, the WHIM is found in filaments, thus representing an important observational tracer of the cosmic web \citep[e.g.,][]{Umehata2019, Cui2019}. Indeed, the baryon fraction of filaments is found to be higher than that of galaxies (however, slightly lower than that of the cosmic mean) and with gas temperatures uniform as a function of the distance to the filament \citep{Gheller2019}. However, unlike the hot gas accumulating in the gravitational wells of clusters, the WHIM is much more challenging to detect. Recent promising results  have lately pushed the study of the cosmic web into the focus of galaxy formation theory, simulation and observations alike \citep{Nicastro, Eckert, Bonamente, Tanimura2019, Tanimura2020, Cui2018b}.
Simulation-based studies have shown that filament properties, such as the average gas temperature, mass, volume and radius, follow scaling relations \citep[e.g.,][]{Gheller2015}.  However, since direct detections often rely on stacking the signal of multiple objects \citep[e.g.,][]{Tanimura2019, Tanimura2020}, a better understanding of the distribution of the WHIM could help improve its detection in future observational studies. 

Cosmological filaments cover a variety of properties ranging from thin tendrils pervading voids to thick bridges between cluster pairs that boast many substructures and collapsed haloes, thus playing an important role of transporting matter into galaxy clusters \citep[e.g.,][]{Cautun, Kraljic2018}. A number of studies using simulations and observations have used cosmic filaments to characterize anisotropic accretion to further establish a picture of steady filament dynamics in less and intermediate dense regions \citep{Pichon2011, Tempel2013, Codis2012, Codis2015,Laigle2015,  Libeskind2015, Kraljic2019}. 
The multiscale and diffuse nature of filaments makes defining and distinguishing between different structures a non-trivial task. To add to the difficulty, different filament-finding algorithms are based on different implicit definitions of these objects. This leads to wide-ranging discrepancies in relation to the nature of identified objects found by the different filament finders \citep{Libeskind, Rost}, including \disperse\ \citep{disperse}, Semita \citep{pereyra}, Nexus \citep{nexus1, Aragon_Calvo2007}, or Bisous \citep{Tempel}. 
Despite these drawbacks, understanding the large-scale structures of the Universe, and how the galaxies they host evolve, remain key science challenges.

Most current research has been focusing on large-scale filaments of dozens of Mpc in length and over large volumes of the Universe \citep[e.g.,][]{Tempel2013, Laigle2015,Codis2015,Malavasi2017,Kraljic2018,Kraljic2019,Veena2019, Cui2018b, Cui2019}.
These investigations have established that cosmological filaments are crucial structures for the dynamics and properties of gas and galaxies alike \citep[e.g.,][]{Codis2012, Cautun}. 
The filaments are surrounded by complex velocity fields that, to a first order, can be explained by the Zel'dovich approximation \citep{Zeldovich1970}. This assumption leads to the expectation of a laminar flux towards the filamentary axis, a manifestation of diffuse material collapsing into filaments along two axes. However, this simple approach does not account for differences between gas and collisionless dark matter or for the dynamics induced by local cosmic web components, like galaxies falling into clusters.

Both observations and simulations have rarely included the special case of filaments around massive galaxy clusters -- dense areas of highly mixed and turbulent environments, where gas undergoes significant shocks \citep{Power2020}.
Likely, the properties of gas and dark matter in cluster filaments therefore differ to that found in filaments in the more general large scale cosmic web. In particular, there is evidence of gas shock processes triggered by the filamentary substructure falling into clusters that could lead to turbulent motion of the gas outside them \citep{Power2020}. Near a filament's spine, cold gas may be expected to flow more smoothly, transporting relatively cold gas into the clusters. This is analogous to the cold flows that are found to shape forming galaxies at high redshifts \citep{Birnboim2003, Dekel2009, Pichon2011, Dubois2012, Danovich2012, Cornuault2018}. These considerations raise questions like: How exactly is gas accreted towards clusters? Is filament accretion more efficient than a purely gravitational dark matter infall?

As the sites of possible pre-processing and matter accretion onto clusters, the study of filaments in cluster outskirts is now gaining momentum as simulations and observations have improved \citep[see][for a recent summary]{Walker2019}. The outskirts of galaxy clusters represent ideal laboratories to study both matter buildup in the Universe and galaxy evolution, as they are convergence points for filamentary networks and offer scenarios of phenomena involving high density nodes, intermediate density filaments and low density "field" regions.
In this paper, we address this specific case and study the gas and dark matter properties of filaments in the vicinity of massive galaxy clusters. We view this analysis of filaments in high density environments as an extension to similar studies that focused on filaments in a wider mix of environments, and on a wider range of scales. 
We investigate the structure and properties of filaments through their {\it diffuse} gas surrounding \halos -- closely related to the WHIM -- as well as the underlying dark matter.

The paper is structured as follows: in Sec. \ref{Sec:data} we detail the simulations of 324 massive cluster our study is based on and the identification of cluster filaments based on gas particles.
In Sec. \ref{Sec:interplay} we investigate the interplay between haloes and filaments and study parameters such as filament length, density, and temperature. 
We also study the role of the clusters that the filaments are connected to and define a characteristic thickness based on density profiles.
In Sec. \ref{Sec:vel_web} we discuss the dynamics of gas and dark matter around the cluster and disentangle flows towards the cluster and towards and along filaments before summarizing our conclusions in Sec. \ref{Sec:conclusions}. In this paper, we are using Planck cosmology with $\Omega_{\rm{M}} = 0.307, \Omega_{\rm{B}} = 0.048, \Omega_{\rm{\Lambda}} = 0.693, h = 0.678,$ \mbox{$\sigma_8 = 0.823$}, $n_s = 0.96$.

\section{Data and filament finding}
\label{Sec:data}
\subsection{\threehundred simulations}
In this work we use $324$ cluster zoom-in simulations at $z=0$ of \threehundred project\footnote{\url{https://the300-project.org}} \citep{Cui2018}. 
The sample was built form a mass-complete sample of clusters by choosing the $324$ most massive virialised objects at $z=0$ within the MultiDark2 simulation volume \citep{multidark}, identified with the \rockstar\ halo finder \citep{rockstar}. The cluster masses range from $M_{200} = 6.08 \times 10^{14} \hMsun$ to $M_{200} = 2.62 \times 10^{15} \hMsun$. Particles in these clusters were then traced back to their initial positions and resimulated using the full-physics code \gadgetx ~ \citep{Springel2005, Murante2010, Beck2016, Rasia2015}, which uses a modern Lagrangian smoothed particle hydrodynamics approach. The high-resolution volumes are spherical regions of radius $15 \hMpc$ centred on each cluster and contain dark matter and gas particles with initial masses of $12.7 \times 10^8 \hMsun $ and $2.36 \times 10^8 \hMsun$ respectively. This usually corresponds to an extent of radius $5R_{200}$ at $z=0$. 
Time steps between redshifts $z \sim 17$ and $z = 0$ are stored in 129 snapshots and are available for a plethora of studies that trace orbits and cluster mass buildup over time \citep{Arthur2019, Haggar2020} 

Haloes and subhaloes were identified in each snapshot using the \ahf
\footnote{\url{http://popia.ft.uam.es/AHF}} halo  finder \citep{AHF}. AHF accounts for dark matter, gas and stars and returns properties that define each halo, such as luminosities, stellar masses, and angular momentum among others. The sizes and masses of halo s used in our analysis refer to $R_{200}$ and $M_{200}$, where $R_{200}$ is the radius of a sphere enclosing an overdensity of 200 times the critical density of the Universe at $z=0$. For this work we only consider \halos\ with masses above $10^{11}$\hMsun (which corresponds to around 80 dark matter particles) and haloes  closer than $15 \hMpc$ from the center of the Lagrangian/re-simulated region, at redshift zero.

For the simulation environment we are studying in this paper, i.e. largely a low overdensity regime, it is important to recognise that Lagrangian particle based methods such as smoothed particle hydrodynamics (SPH) are not well suited to recover sharp discontinuities such as shocks due to the enforced relatively long search lengths required to find a suitable number of particle neighbours. Adaptive mesh or moving mesh based methods would be better able to resolve these. As such we need to be careful with our analysis and bear this in mind when discussing our results. Effectively, small scale structures cannot be readily resolved within the low density material. For this reason we generally stack our data, a process which would naturally smooth these structures anyway. Further, we do not employ the recovered SPH gas density from \gadgetx\ but rather recalculate density from the stacked particle locations. Lowering our particle mass and so improving our ability to resolve sharp discontinuities would improve our ability to see small structures but any such effort is limited because a factor of two decrease in linear scale requires nearly an order of magnitude increase in particle number. We note that the size of the structures discussed below is significantly above our resolution limit in this sense and as such they are well resolved.

\subsection{Filament extraction based on gas particles}
\label{sec:filament_extraction}
\begin{figure}
\centering
\includegraphics[scale=0.6]{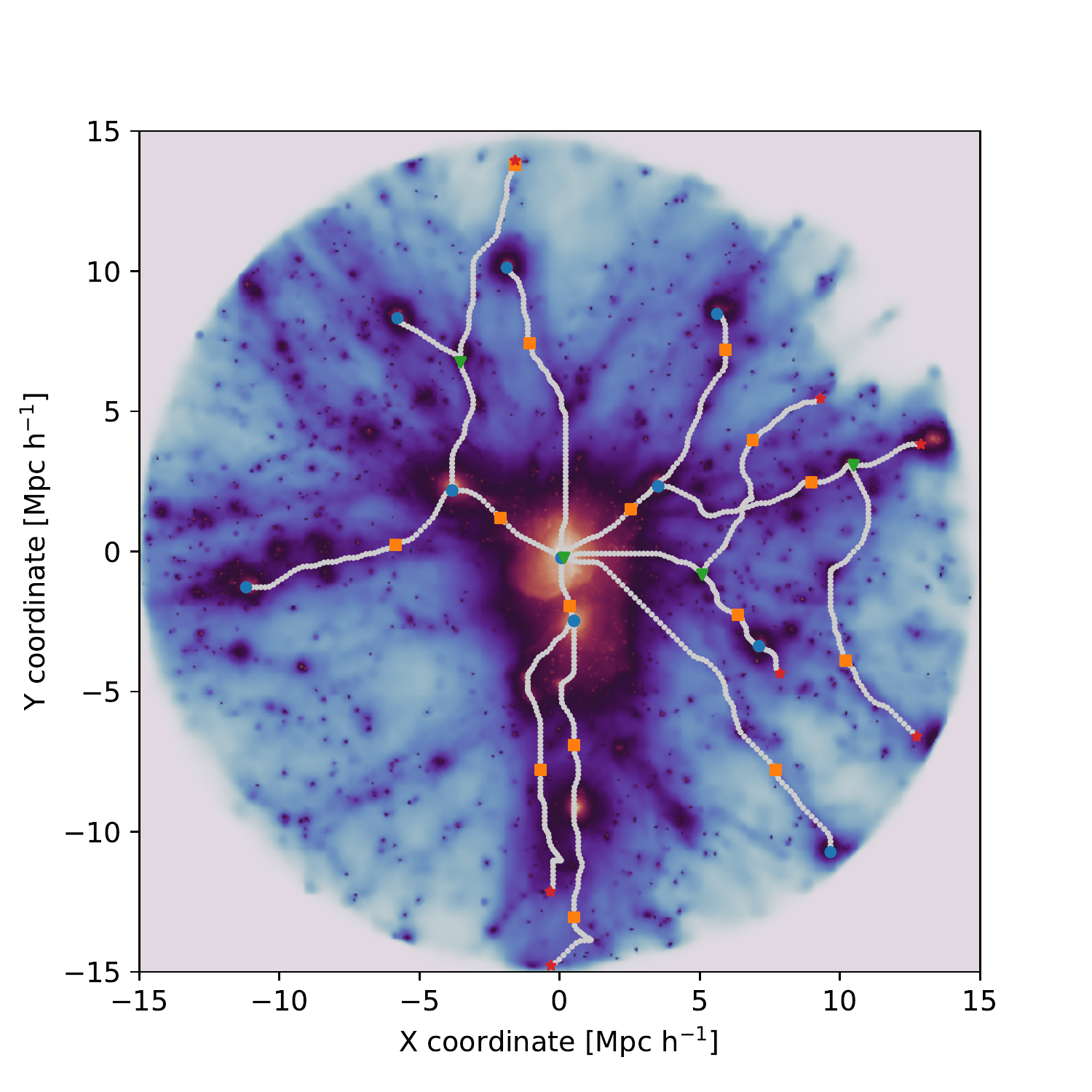}
\caption{One example cluster of \threehundred simulation suite with its extracted filament network based on gas particles. The background shading indicates projected gas particle density along the $z$ axis of the cluster, rendered with SPH-viewer \citep{Benitez_Llambay2015}. The recovered \disperse\ filamentary network, which was identified using the 3D distribution, is shown in white, symbols indicate different features of the filamentary network. Blue circles: nodes, orange squares: saddle points, inverted green triangles: bifurcations and red stars: exit points.}
\label{fig:data}
\end{figure}

We identify filaments with the filament finder \disperse\ \citep{disperse}, a tool designed to extract multi-scale structures. From a given distribution of tracers, \disperse\ identifies significant topological features in the density field (maxima, minima and saddle points) as well as ridges that connect them. 
In short, \disperse\ provides the positions and identity of a number of web-like structural features where the filaments themselves are the lines connecting the singularities, defined by the gradient field \citep{disperse, Libeskind}. In this context {\it nodes} can be thought as an extreme point where DM haloes  reside and {\it saddle points} are next extreme point along the filament axis.

For our analysis, we extract filaments based on the 3-dimensional number density of gas particles in each simulation box. The gas density was first binned in a 3-dimensional grid of $30\hMpc$ width and pixel resolution of $150~\rm{kpc ~ h^{-1}}$ on each side. This grid was then smoothed over eight pixels with a gaussian kernel. This process allows us to focus on large cosmic filaments that connect groups and cluster centres, and excludes thin tendrils. \disperse\ then identified filaments using an absolute persistence cut of 0.2 (corresponding roughly to a $5\sigma$ persistence threshold using halos). The persistence is defined as the ratio of the density value of a pair of critical points (e.g., node and saddle point) and governs the robustness of the feature and is used to additionally filter out low significant filaments.

\disperse\ returns the identified ridges as a set of segments that construct the persistent skeleton of the network as well as the topologically robust extrema, i.e., saddle points and nodes. Figure \ref{fig:data} shows the filament network of Cluster 1 (plotted in white on top of the gas density in blue shading) as an example. In addition to nodes (blue points) and saddle points (orange squares), we also show bifurcation points that mark the location where a filament splits (green triangles). Often, bifurcation points indicate local peaks in the density field, however they are deemed less significant than nodes. Red stars indicate exit points: this is the location where a filament leaves the extent of the simulated zoom region around the cluster. Combining all networks in 324 zoom regions, we thus identify a sample of $11,058$ filaments based on gas particles.  
While the example shown in Fig. \ref{fig:data} offers an obvious clue that the identified network indeed traces the underlying structure of the density field, we refer the reader to \citet{Kuchner2020}, in which detailed tests of the filament extraction method on gas and mock galaxies in the same simulations, as well as comparisons between detections in 2D and 3D are presented.

In this work we define a filament as a structure that connects a topological peak (node) with a saddle point, sometimes splitting along the way into multiple branches at bifurcation points. Thus, any single node \textit{controls} all filaments connected to that node as far as the first saddle points. Naturally, filaments continue beyond the saddle points, but they are then controlled by a different node. As a basic principle, material in filaments on either side of the saddle point will eventually fall into a different halo. This can be thought of as similar to the way water on either side of a watershed may end up in different oceans. Note that this physically motivated definition of filaments may differ from other definitions in the literature that often define a filament as joining two nodes. Several filaments together constitute a filament network which is the basis for our analysis.
\begin{figure}
\centering
\includegraphics[width=\columnwidth]{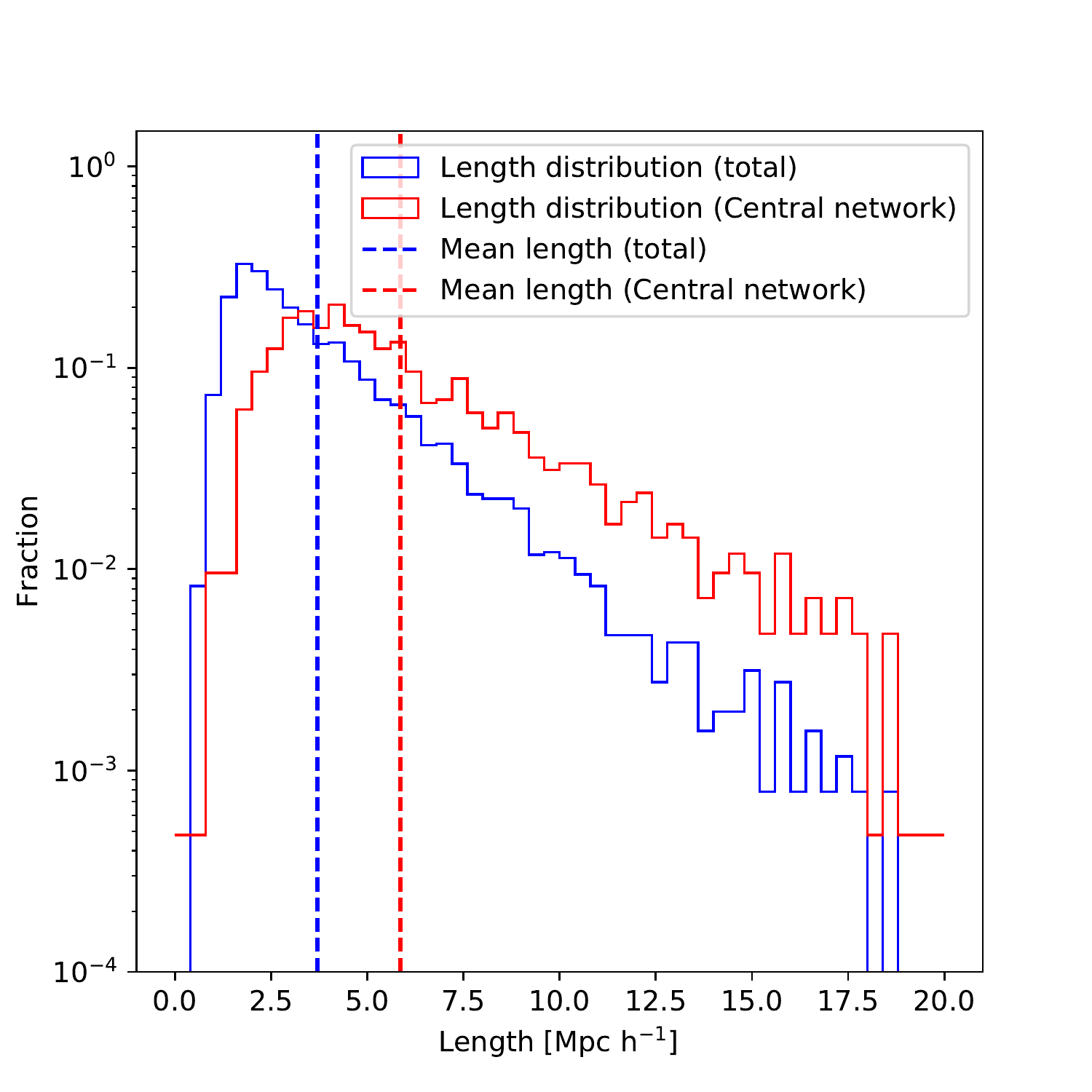}
\caption{The length distribution of filaments in our sample of 324 clusters is longer for filaments associated with central nodes. Only filaments with a node and a saddle point were considered, those with bifurcation or exit points were excluded. The blue distribution shows that the average length is approximately $3.70$\hMpc\ for the total sample and $5.88$\hMpc\ for filaments connected to the central cluster (shown in red). Filaments run from a node to a saddle point and at least two are needed to bridge between \halos.}
\label{fig:hist_length}
\end{figure}

\section{The interplay between haloes and filaments}
\label{Sec:interplay}
Filaments are regions of moderate collapse hosting both diffuse material and already bound haloes. These components interact dynamically with one another through accretion, outflows and stripping. To some extent halo  gas, intra-halo  gas and filament gas mix, thus making the study of the pure diffuse filament gas a challenging task.
We approach this challenge by isolating the gas in filaments from the contribution of embedded haloes.
This is not easy in the best of cases and it is important to note that this exercise is complicated by resolution: every simulation's implementation of gas density, temperature and cooling model governs how well pure spine flows and haloe-outskirt particles can be separated. The exclusion of the halo  contribution therefore always represents a compromise.

In this section, we first look at the contribution of the main halo\footnote{In \threehundred simulations, the mass of the central halo  in each simulation box is comparable to the entire cluster's mass itself and therefore influences the surrounding cosmic web significantly.} on the lengths of filaments connected to this halo  (Sec. \ref{subsec:filament_lengths}), before examining the removal of haloes in filaments -- which can represent the main cluster, groups and individual galaxies -- in more depth (Sec. \ref{subsec:haloes_filaments}). Finally, we determine (Sec. \ref{Quantities}) and characterise (Sec. \ref{sec:filament_thickness}) filament profiles based on gas and dark matter densities and contrast the contributions of nodes and distances to the nodes and filaments on these profiles (Sec. \ref{subsec:combined_effect}).

\subsection{The dependence of filament length on the proximity to the central node}
\label{subsec:filament_lengths}
\begin{figure}
\centering
\includegraphics[scale=0.6]{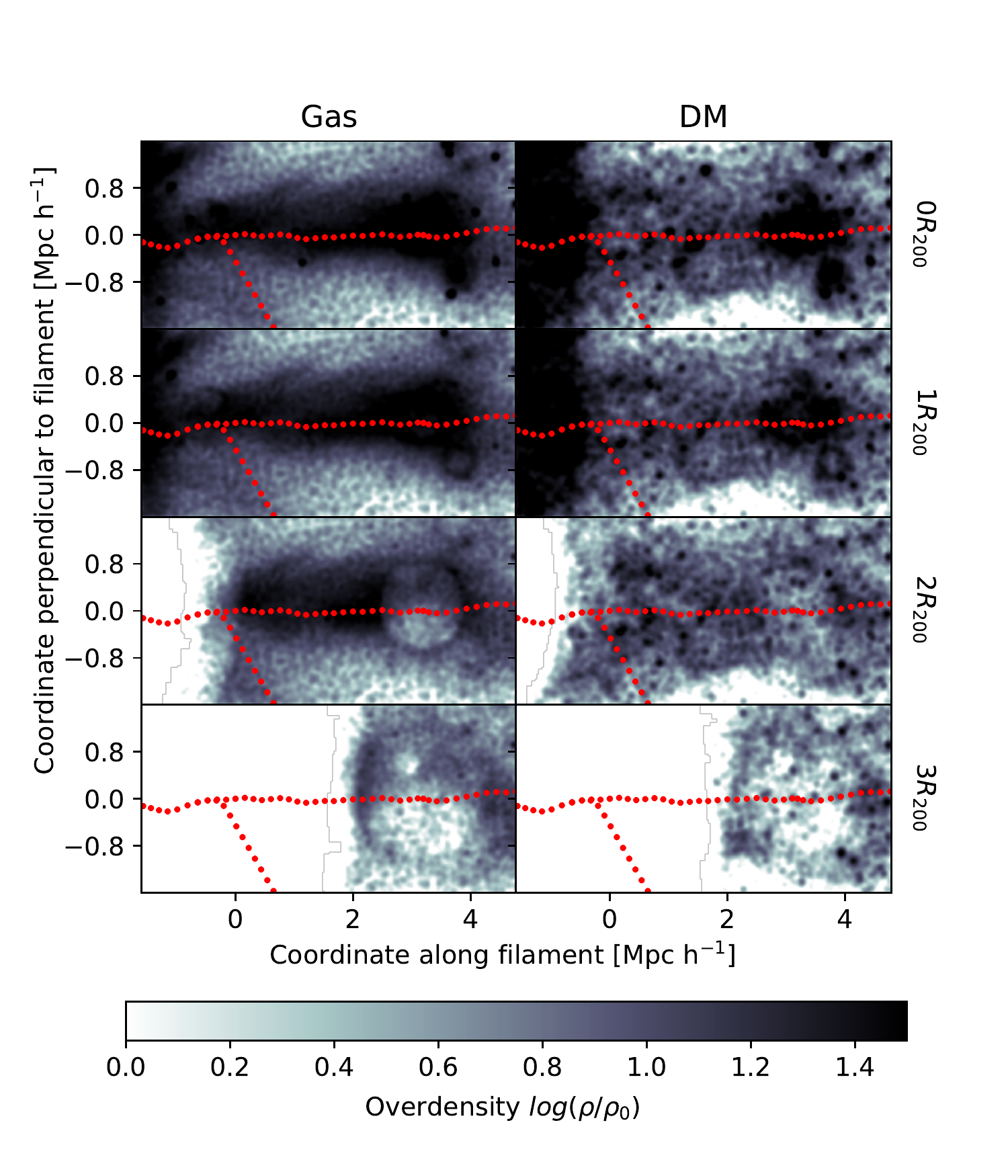}
\caption{Illustration of the process of isolating diffuse filament material by removing increasingly larger regions around haloes: Column-integrated overdensity of regions surrounding filament 4 of cluster 2. From top to bottom, overdensities after removing $0$, $1$, $2$, and $3$ times $R_{200}$, respectively. On the left: gas particles. On the right: dark matter particles. \disperse\ filaments are marked in red, note that the central cluster is located to the left side.}
\label{fig:vol_render1}
\end{figure}
Processes that give rise to filament properties are to a high degree scale-invariant leading to the multi-scale nature of filaments. The choices we make during the filament extraction therefore governs the range of scales we are able to investigate. In this study, we choose to focus on cluster filaments, i.e., prominent filaments in outskirts of clusters with $M_*>10^{14} \rm{\Msun ~ h^{-1}}$, and have adapted the background density and smoothing accordingly.

To introduce the scales of filaments we are studying in this paper, Fig.~\ref{fig:hist_length} shows a histogram of the lengths of all recovered filaments. These filaments have a wide range of lengths, with a small number extending from the centre of the cluster all the way to the edge of the re-simulated region.
Naturally, inter-group filaments are on average shorter than inter-cluster filaments. 
We distinguish between filaments that are connected to the central halo  (shown in red) which we call the central network, and all other filaments. 
For the present case of \threehundred simulations, filaments connected to the central cluster are around $6$\hMpc\ long while those originating anywhere in the volume are on average about half this length. Of course, these numbers are influenced by the limited volume we are probing in this study. 

Other studies such as \citet{Malavasi2020, Tanimura2020, Gheller2015, Gheller2019} focus on long filaments of dozens of Mpc, whereas others like \citet{Kraljic2019} focus on filaments in the range $3-12 \rm{Mpc}$.  
Here, we report the distance between a node and a saddle point, the length of filaments from one node to the other will typically be twice this value as normally both halves of a bridge between two nodes will be counted separately. 
Although the length range is not comparable to other studies mentioned above, the trend of having a lower population for longer filaments seen here is widely observed in the literature \citep{Rost, Galarraga_Espinosa2020, Malavasi2020, Bond2010}.
Note also that the recovered length and number of filaments depends on the persistence level chosen during filament extraction, with a lower level leading to more nodes and a finer network. As mentioned in Sec. \ref{sec:filament_extraction}, we have deliberately chosen a procedure to robustly extract the main branches of the network that omits small tendrils.

\subsection{The contribution of haloes in filaments}
\label{subsec:haloes_filaments}

Haloes are an integral part of filaments. Our aim in this paper is to investigate filament profiles based on the diffuse material of filaments, i.e., without the contribution of haloes. To achieve this, we remove the gas that can be attributed to haloes.
A precise and complete removal is, however, unattainable due to the impossibility of clearly determine the boundaries of halos. We therefore approximate the process by removing material inside spheres around haloes -- detailed in this section -- leaving only {\it diffuse} material. 
%
\begin{figure}
\centering
\includegraphics[width=\columnwidth]{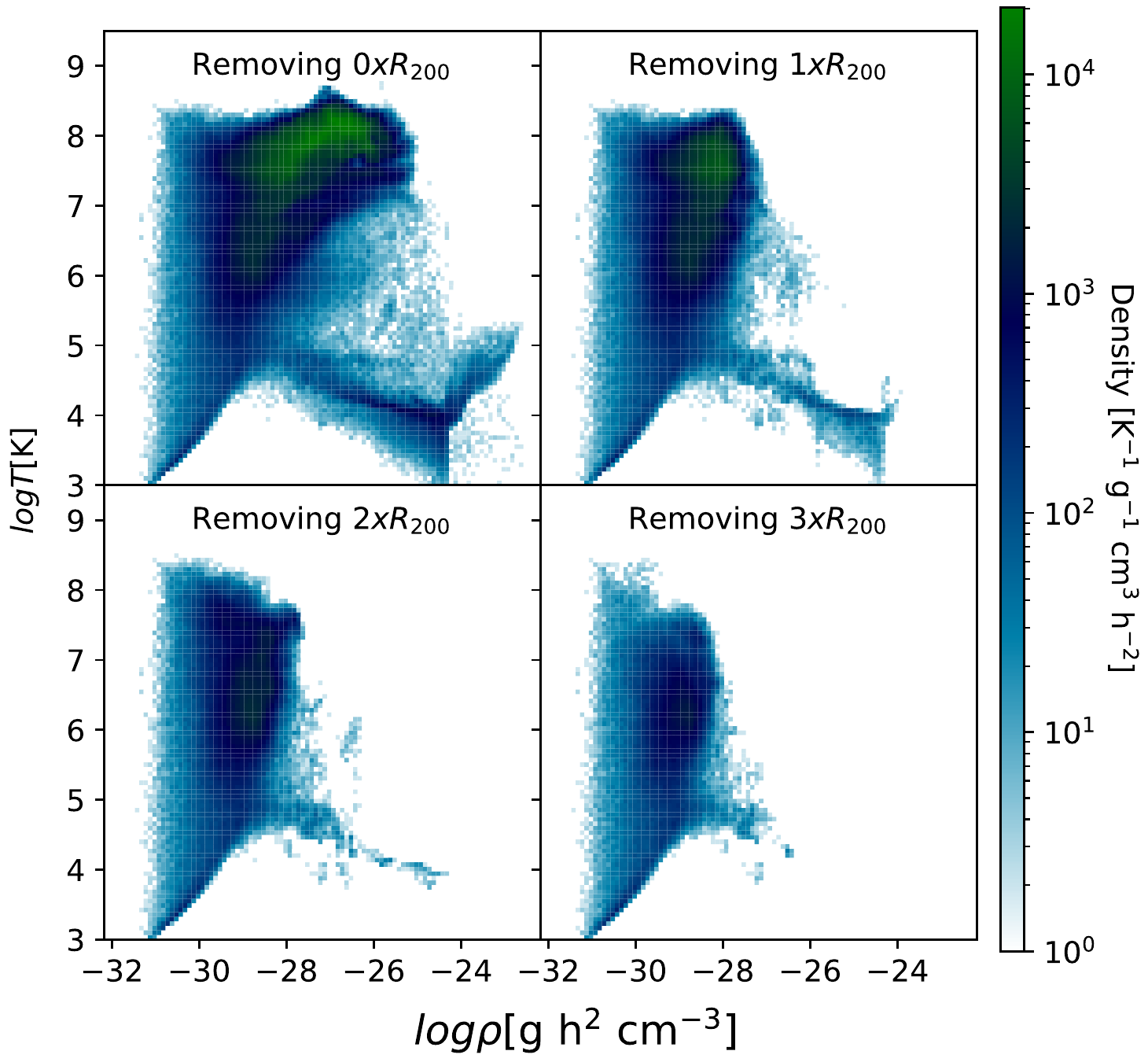}
\caption{Demonstration of the contribution of haloes on the density -- temperature phase-space diagram of gas particles contained within the filaments of cluster 0001. The top left panel displays all the particles while in the other panels successively larger regions around each halo  of which mass limit, $10^{11}$ are excised as indicated, the colour represents the value of the distribution function.}
\label{fig:factor_comp2}
\end{figure}

\begin{figure}
\centering
\includegraphics[scale=0.6]{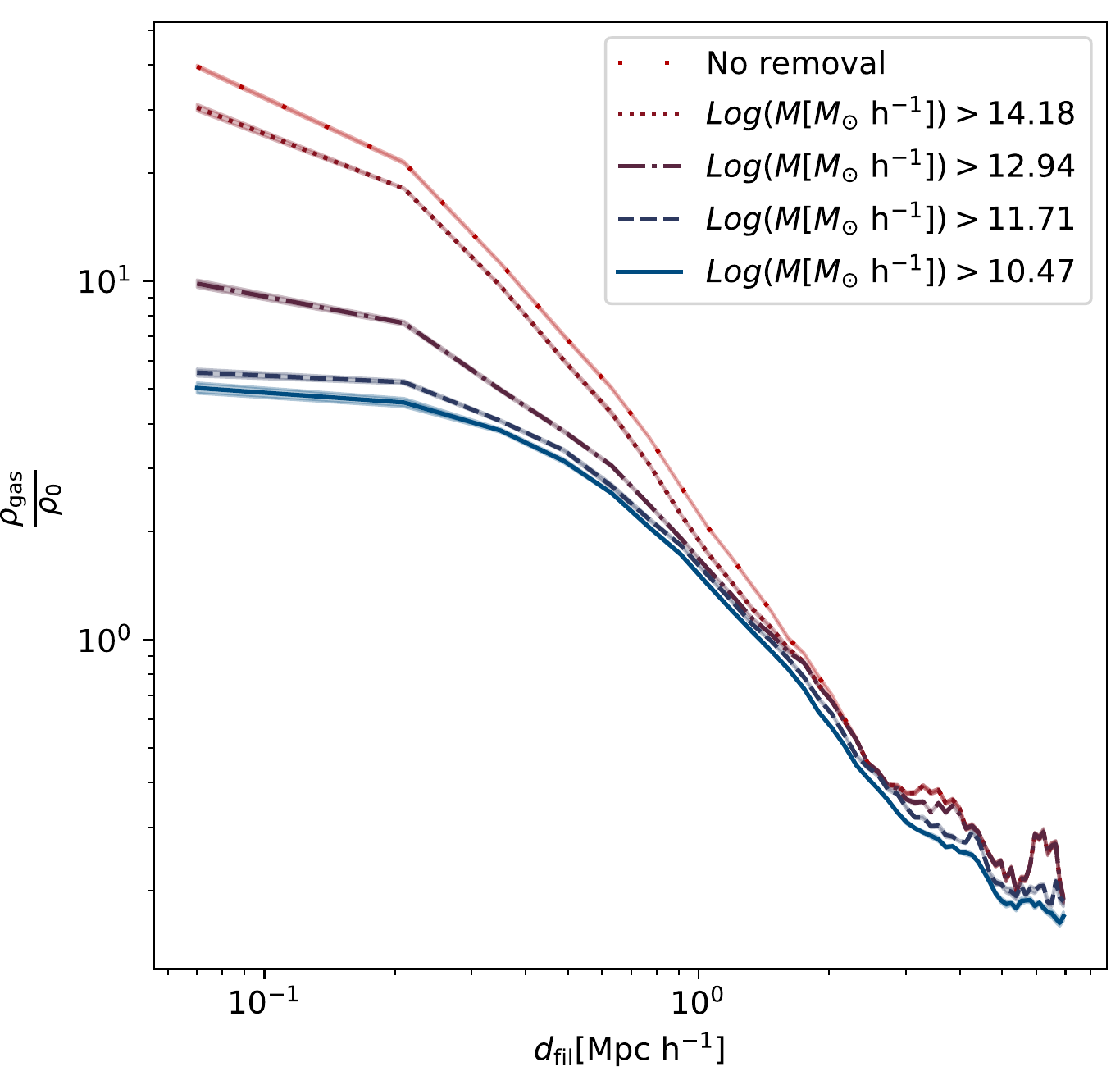}
\caption{Filament profile and the contribution of haloes: Overdensity profile of gas particles as a function of the radial distance to the nearest filament axis for filaments from the largest cluster. The shape of filament profiles depends on the contribution from haloes, becoming steeper as haloes with progressively higher mass limits are removed. The particles within $2 \times R_{200}$ of \halos\ with a certain indicated minimum dark matter mass are excluded.}
\label{fig:profile_linear}
\end{figure}

First, we need to find the appropriate size of the region of halo  influence. To do so, we exclude particles within increasingly larger regions around increasingly more massive haloes. Specifically, in this test we exclude particles within 0, 1, 2 and 3 times $R_{200}$ (see Fig. {\ref{fig:vol_render1}} and \ref{fig:factor_comp2}) and around haloes with log halo  masses greater than 10.47, 11.71, 12.94 and 14.18 $\hMsun$ (see Fig. {\ref{fig:profile_linear}}). We first detail the effect of varying sizes of regions before looking at identifying the appropriate halo  mass limit.

To illustrate the removal process, Figure~\ref{fig:vol_render1} shows a volume rendered example filament in both gas and dark matter with successively larger excision radii from top to bottom. Removing a volume of $1 \times R_{200}$ around each halo  leaves a high density shell around the centres of some haloes. By removing particles within $3 \times R_{200}$ of haloes, we risk excluding filament gas. Thus, exciting volumes equivalent to $2\times R_{200}$ seems like the best compromise.

Figure ~\ref{fig:factor_comp2} shows an alternative view of the contribution of halo  gas and the removal process. The figure shows the density - temperature phase space diagram for all the particles found within the central $10\hMpc$ of the largest cluster in the sample (the same as shown in Fig.~\ref{fig:data}). The diffuse material of filaments is highly related to the Warm Hot Intergalactic Medium (WHIM), as its gas is expected to be distributed along the cosmic web, surrounding galaxies embedded in filaments. 
Fig.~\ref{fig:data}) highlights the thermodynamical properties that characterize both haloes and the diffuse filament material.  
In the four panels, we progressively remove larger volumes surrounding each halo  as before. Some of the particles lie within the central hot cluster halo  with temperatures around $10^8$K -- these identify the diffuse intra-cluster medium -- while others are gas at high overdensity and temperatures close to $10^4$K, these are the multi-phase particles with a significant cold fraction which represents the reservoir for the star forming process. As successively larger regions around each of the identified \halos\ are excised (1, 2 and 3 times $R_{200}$ respectively as indicated in the panels), we see both this hot halo  gas and the galactic phase largely disappearing. 
Removing particles out to $1\times R_{200}$ (top right panel in Fig.~\ref{fig:factor_comp2}) leaves a considerable amount of warm-diffuse particles. By increasing the limiting radius in units of $R_{200}$, the maximum density decreases significantly and both the hot and star-forming phases are largely excluded. The bottom two panels indicate that essentially only underdense material and WHIM remain once regions inside twice $R_{200}$ are removed. 
After removing the effect of haloes, the temperature and density distributions remain stable along the filament spine. We only see a small increase of the gas temperature closer to nodes, while the gas density is largely unchanged. Note that the filaments were identified before the removal process.

We also investigate the impact of haloes of different masses. This is highlighted in Fig. \ref{fig:profile_linear}, which shows the profiles of gas filaments after removing volumes with radii of $2 \times R_{200}$ surrounding progressively more massive \halos\, as indicated in the caption. We explain the procedure of obtaining profiles in the next subsection.  
The largest halo  in our volume has a mass of $10^{15.41} \hMsun$, representing the main cluster itself. Massive haloes constitute the majority of the filament density (top lines in Fig.~\ref{fig:profile_linear}). This levels off at lower masses.
Since removing haloes above a mass of $10^{10.47}\hMsun$ and above a mass of $10^{11.71}\hMsun$ (the two lower profiles) does not substantially modify the filament-perpendicular profile, we decide to remove all haloes with masses above $10^{11}\hMsun$ for the removal process. In this process, due the resolution limit of our simulations, we know that we will work with environments that may lack of physics that is not fully resolved (i.e. turbulence). However, as described in the next sections, we calculate general properties (by averaging or stacking) that will erase the small scale fluctuation making our results robust.

In summary, we conclude that removing volumes equivalent to $2 \times R_{200}$ around haloes with halo  masses larger than $10^{11}\hMsun$ largely uncovers the signal of the filaments themselves, and it is this value that we use for the remainder of this paper. 
While we continue our analysis by reviewing filament material after removing particles in these regions, we acknowledge that this is a choice that comes with drawbacks. We expect some degree of gas cooling and increased density due to shocks of the spine of filaments \citep{Dekel2009} which in turn helps to build up haloes and galaxies more efficiently near the spine. This leads to a correlation between halo  and filament positions which adds a challenge to distinguishing between pure spine flows and halo  outskirt particles. These cooler flows are an integral part of the filament and by excising massive haloes, we inevitably also remove some filament gas along the spine.

\begin{figure}
\centering
\includegraphics[scale=0.4]{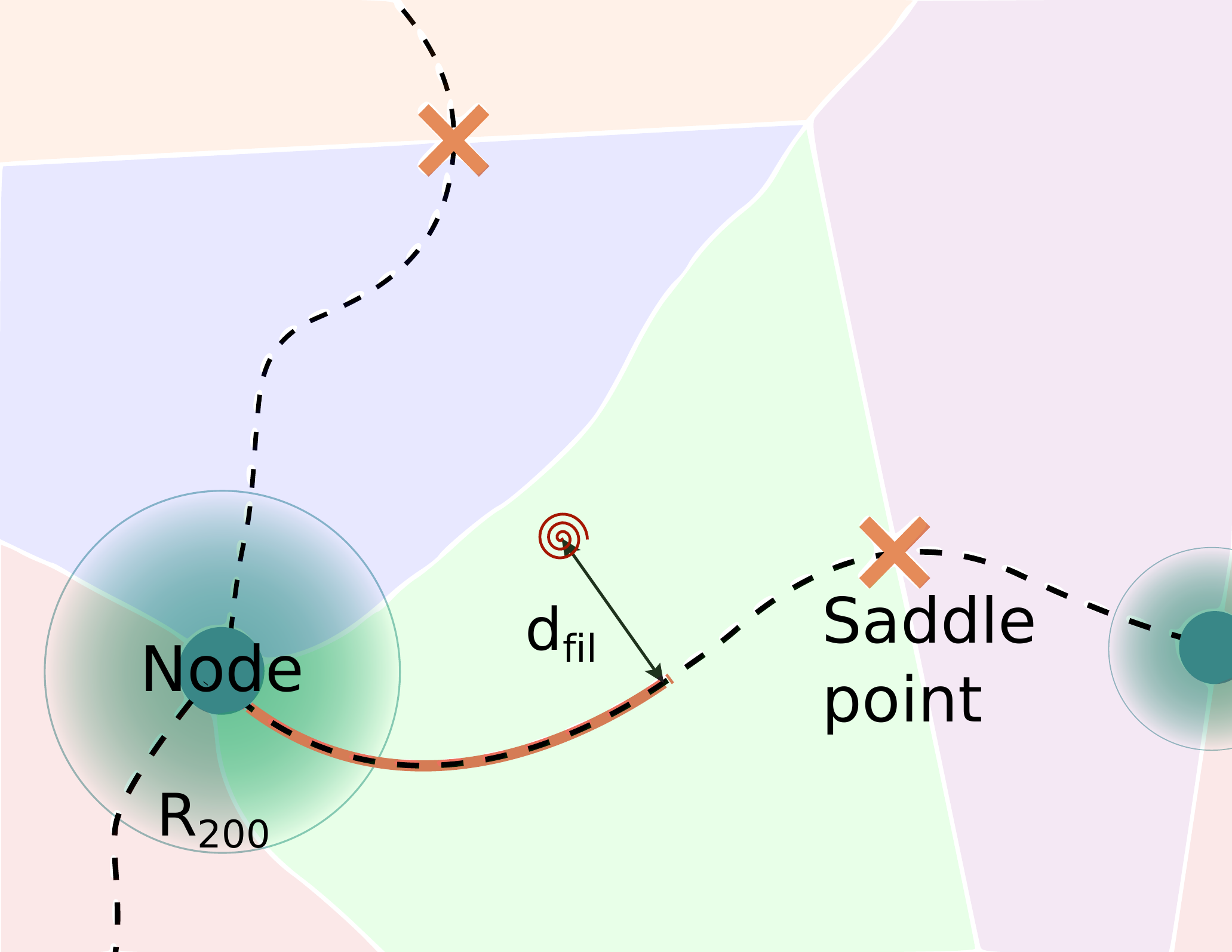}
\caption{Diagram of the quantities used in this work:
Illustration of a section of a network displaying nodes with their $R_{200}$ spherical region,  saddle points -- marked with 'x' -- and the filaments as dashed lines. An object's $d_{\rm fil}$ is defined as the perpendicular distance to the closest filament, whereas $d_{\rm node}$ is defined as the distance to the node along the filament (indicated by the orange line). We present these values normalised by $R_{200}$.}
\label{fig:diagram}
\end{figure}

\subsection{Filament density profile determination}
\label{Quantities}

We first define important quantities (see Fig. \ref{fig:diagram}) needed for the determination of filament profiles.
One such quantity is $d_{\rm{fil}}$, the perpendicular distance of every gas and dark matter particle to the nearest filament segment\footnote{Filaments are described by \disperse\ as a skeleton made up of a succession of short segments.}. Note that particles whose closest segment is the final segment (filament end) are dropped from the analysis as these particles do not have a perpendicular distance to the filament itself, lying beyond the end of the structure. 
This approach is in line with other studies, including work by \citet{Kraljic2018, Malavasi2017} and \citet{Tanimura2020}. It has the advantage of automatically following curved filaments over the 
alternative approach of imposing a cylindrical region of interest \citep[adopted by, e.g.,][]{Colberg, Martinez, Tanimura2019}.
In addition, this means that only the nearest filament to each particle is considered, avoiding signal contamination from other filaments. 

In order to statistically study the influence of nodes on gas and dark matter in filaments, we stack the signal across all simulations. To do so, 
we need to 1) characterise node properties and 2) calculate and normalise the distance of any given particle to the node along the spine of the filament (see Fig. \ref{fig:diagram}). 
Characteristic quantities like node size or mass are not identified by the \disperse\ algorithm. 
As an approximation, we find all \halos\ within a sphere of radius $0.4 \hMpc$ centred on each of the node's position as retrievd by \disperse. The properties of the most massive halo in this area act as a proxy for the node. In this way we can separate nodes into small, medium and large nodes in the following sections. 
We further define $d_{\rm{node}}$ as the distance along the filament of a particle's closest segment to the node (Fig. \ref{fig:diagram}), normalised by $R_{200}$ of the halo  node. Note that $d_{\rm{node}}$ does not allow us to easily study saddle points, as in terms of $d_{\rm{node}}$, the positions of the saddle structure will vary from each filament. 

\medskip

The perpendicular filament density profiles in this paper are presented separately for gas and dark matter particles. We calculate them by dividing the count of particles in bins of distance to the filament ($N_{\rm{p}}(r_{\rm{i}}, r_{\rm{i + 1}})$, for p particles), and stacking over numerous populations of filaments. We then normalise by the total volume occupied by the particles in each bin. This is estimated by using a set of $N_{0} = 5000000$ randomly positioned particles occupying the same spherical volume of radius 10 Mpc/h and determining the number count in the same distance bins $N_{\rm{random}}(r_{\rm{i}}, r_{\rm{i + 1}})$.

Then, the over-density profile of each type of particle p can be defined as:

\begin{equation}
\label{eq:1}
\rho_{\rm{p}}/\rho_0 = \frac{ N_{\rm{p}}(r_{\rm{i}}, r_{\rm{i + 1}})} {N_{\rm{random}}(r_{\rm{i}}, r_{\rm{i + 1}}) }   \frac{ N_{0} m_{\rm{p}}}{V_{0}  \rho_0},
\end{equation}

where $N_{0}$ is the amount of random particles in the total volume $V_{0}$, $\rho_0$ is the average cosmic density and $m_{\rm{p}}$ is the mass of the particle. When applied to all the particles in the full volume this formula gives us the over-density of the filament, but it includes a high degree of contamination from the enclosed \halos.

The multiplicity of filaments, their faint signal compared to high density structures, and the added difficulty as a result of the removal of haloes (Sec. \ref{Sec:interplay}) challenges the study of individual filaments.
We thus infer the statistical properties of populations of filament (e.g., a set of long filaments, or all filaments connected to the centre of a cluster) by employing the {\it stacking} method.
With this, all filaments in the sample contribute to the signal that we measure, thereby improving the signal-to-noise ratio and removing small scale fluctuations that we may not be well resolving by lack of resolution.

For results presented in the following sections, we have therefore combined the particles of all $324$ cluster realisations, and separated by different populations of filaments (e.g., short and long filaments, as explained in the following). 
The high number of filaments ($11,058$) ensures meaningful statistical samples.
We estimate the standard deviations of the density/fraction curves with the {\it Jackknife} method. We build $324$ samples of $323$ clusters each, excluding the i-th cluster, and recalculate the profiles/fractions. In contrast with rather building samples of filaments,  with this method we allow the cosmic variance to contribute to the error of the signal, since the total cluster masses differ significantly between each other.

\subsection{Dark matter and gas characteristics of cluster filaments}
\label{sec:filament_thickness}

\begin{figure}
\centering
\includegraphics[width=\columnwidth]{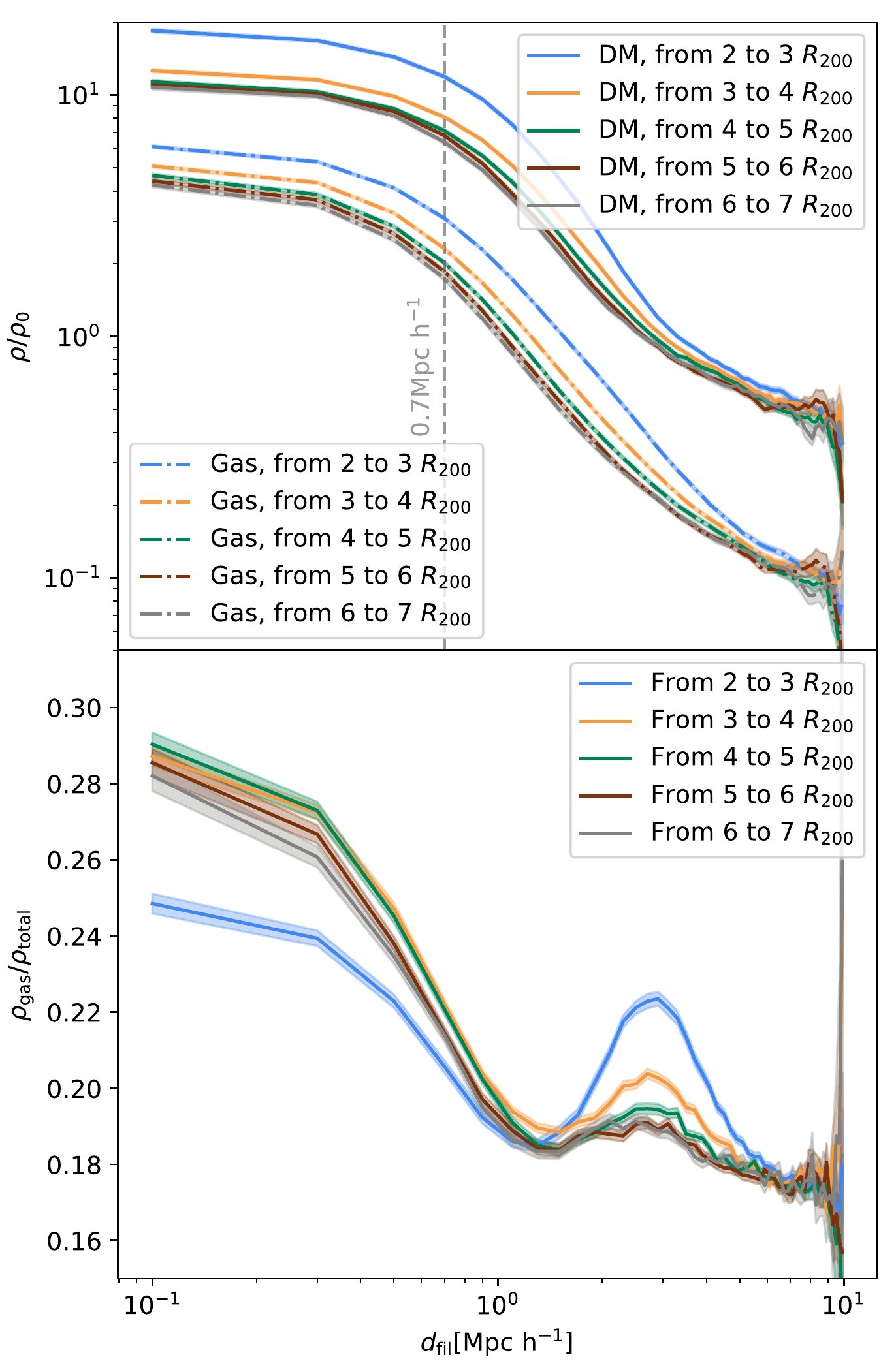}
\caption{Filaments are thicker closer to nodes. Top: Stack of density profiles perpendicular to the filaments for gas and dark matter particles normalised by the cosmic average density $\rho_{0}$. Solid lines indicate the dark matter profile, dashed lines the gas profiles. Various colours differentiate distance from the filament node. Bottom: gas fraction, determined as the ratio between the gas profile and the total DM + gas profiles. Only filaments with nodes and saddle points and without bifurcation or exit points were stacked. Particles within $2$ $\times R_{200}$ of \halos\ above a minimum mass of $10^{11}\hMsun$ are not considered (see text for details). The estimated standard deviation of the curves is shown as a shade of the same curve's colour.}
\label{fig:overdensity}
\end{figure}

The density profiles allows us to quantify a characteristic filament "thickness" at local haloe-scales. However, filaments are not uniform and their thickness may vary depending on proximity to nodes and depend on the used mass proxy, either gas or dark matter. In the following section, we investigate these variations. 
In our scenario, nodes typically mark the positions of large haloes akin to galaxy clusters and groups, towards which matter is flowing via inter-cluster filaments. In Fig.~\ref{fig:overdensity}, we split particles (both dark matter and gas) into six bins of $d_{\rm node}$: $[2, 3]$, $[3, 4]$, $[4, 5]$, $[5, 6]$ and $[6, 7]$, indicated by lines of different colours. Dot-dashed lines show profiles for gas particles and solid lines for dark matter particles. 

The top panel in Fig.~\ref{fig:overdensity} reveals a "cored" radial density profile up to approximately $0.7\hMpc$ distance to the filament axis, as marked in the figure. We use this value as a "typical thickness" for cluster-filaments as revealed by \textsc{The ThreeHundred}. At larger distances, the profiles follow a power law $r^{-\gamma}$, with $\gamma \approx 2$. A similar shape is also seen in other works with different emphases and starting points, e.g. \citet{Colberg} who used $\Lambda CDM$ numerical simulations to study the filaments between pairs of clusters, and \citet{Dolag} who studied gas density profiles in hydrodynamical resimulations \citep[see also][]{Bonjean2019, Galarraga_Espinosa2020, Aragon_Calvo2010}. Note, however that the shape of the gas density profile is resolution and cooling-model dependent.

It is not trivial to define a typical "thickness" of filaments, let alone compare between different studies, since this depends on individual definitions. In the literature, some studies determine the radius of a filament as the distance from which the density profile starts to follow a power law \citep[e.g.,][]{Colberg, Aragon_Calvo2010}. They find a value for a characteristic filament radius of $\approx 2 \hMpc$. 
Others define the distance at which the density drops below a certain threshold \citep{Dolag},which leads to higher values of $5\hMpc$ radius. 
Yet others use the scale parameters that fit the data using a certain model, finding thickness parameter values of $r_{2} = 2-3 \hMpc$, with long filaments being thinner than short filaments \citep{Galarraga_Espinosa2020}, 
or $r_{m} \approx 7.5$~Mpc for a exponential decay fit for long filaments $>20$~Mpc \citep{Bonjean2019}.

\begin{figure*}
    \centering
    \includegraphics[width=1\linewidth, keepaspectratio]{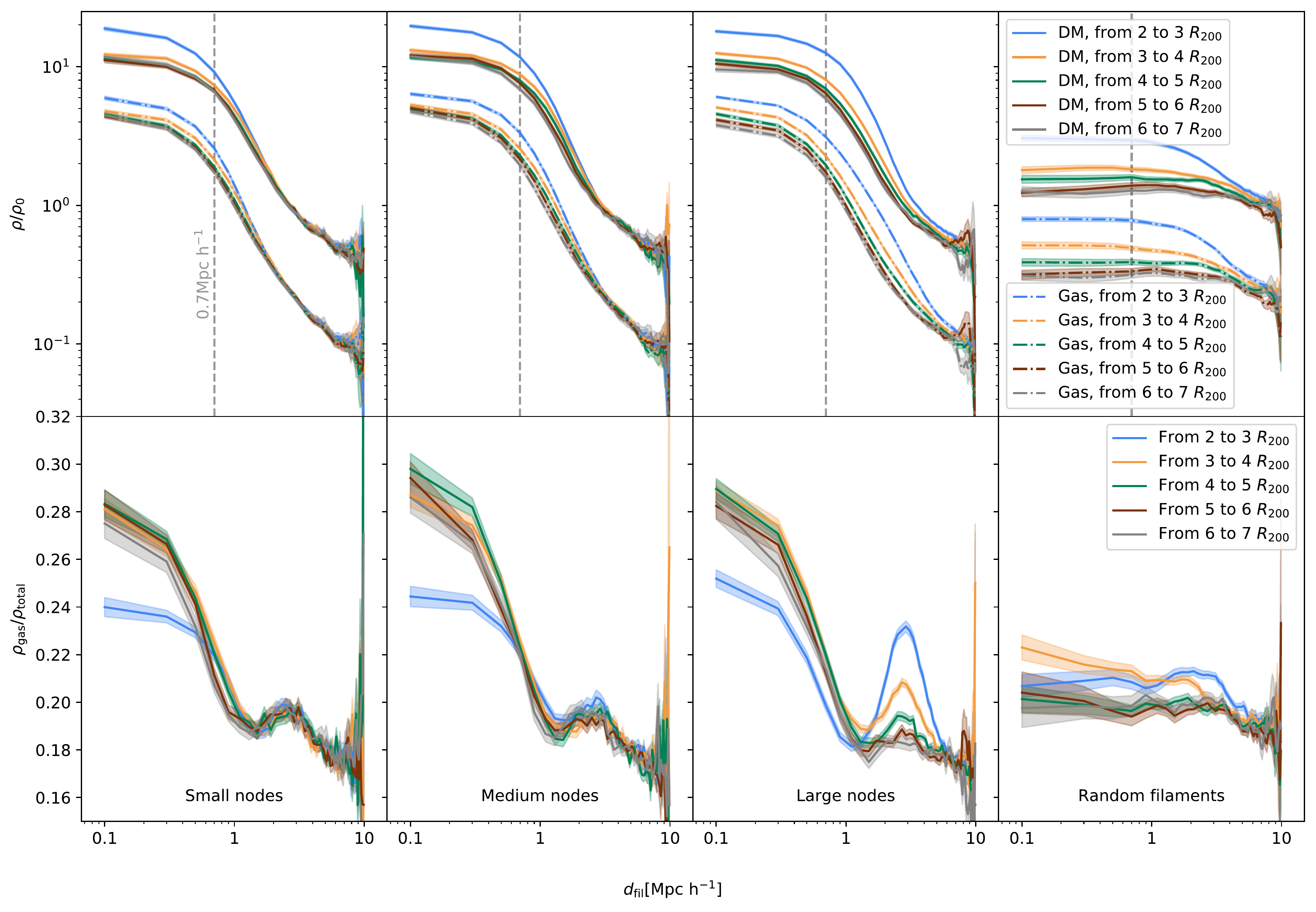}
    \caption{Combined effect of haloes and distance to nodes: Perpendicular density profiles for stacked filaments. As for \ref{fig:overdensity} but split by node size. Left column displays the smallest nodes while the third column displays the largest nodes, usually the central \halos\ of each simulation. The last column shows the density profiles of filaments that had been randomly rotated about the node (see text for details).}
\label{fig:density_splitbysize}
\end{figure*}

For gas and dark matter alike, the peak overdensity rises towards the nodes by a factor of around two (indicated by different coloured lines). This could be due to the remaining influence of the central cluster which extends beyond the radius we use to remove its contribution ($2 \times R_{200}$). The thicker filaments closer to nodes could therefore be related to halo  outskirt particles that mix with filament particles, or rather be a genuine correlation of the filamentary structure and the proximity to a node. This effect disappears beyond three times $R_{200}$. This observations is also seen in other studies, e.g., \citet{Dolag}, who also find that gas profiles increase to higher values closer to nodes. Furthermore, the power law $r^{-\gamma}$ that profiles follow changes from $\gamma \approx 3$ near the node to $\gamma \approx 2$ further away. This is consistent with a divergence from an NFW profile at the node \citet{NFW}. The varying peak density and width are also seen in \citet{Kraljic2019}, with higher densities towards the nodes in the galaxy distribution and lower width of the isocontours at the saddle point.

Fig. \ref{fig:overdensity} also shows that local properties of the filament profiles from gas particles (dot-dashed lines) differ from those based on dark matter particles (solid lines), where dark matter filaments are somewhat thicker, i.e., they appear less concentrated than gas filaments. This is especially noticeable for bins $[2, 3]$ and $[3, 4]$  $R_{200}$ distance from the node, where the drop is noticeably steeper for dark matter particles. This could imply that the gas around \halos\ is more homogeneous than dark matter or that dark matter particles depend more strongly on the distance to the node. 
Alternatively, gas filaments around \halos\ could be more concentrated due to the ability of the gas to shock and cool while dark matter is collisionless (see also Fig. \ref{fig:vol_render1} in which the dark matter filaments appear "fluffier"). This can have important implications for the galaxies themselves \citep{Birnboim2003, Pichon2011, Danovich2012} and their distribution in clusters \citep{Welker2020}. Note that this result refers to massive clusters at z=0, where gas filaments could be more perturbed than at higher redshifts \citep{Dubois2012}.
\begin{figure}
\centering
\includegraphics[width=\columnwidth]{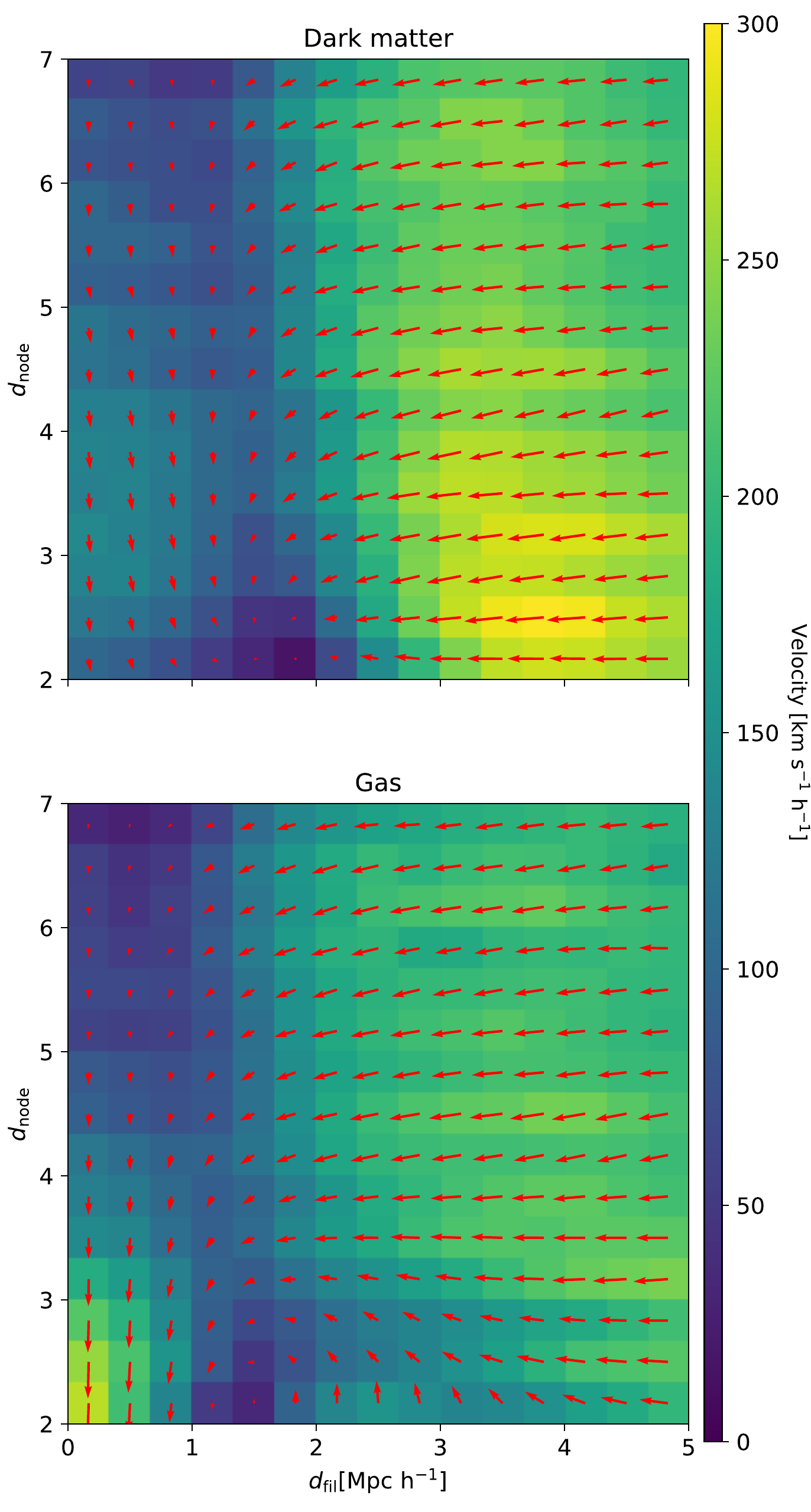}
\caption{Dynamics of filaments: Velocity field of dark matter (on the top) and gas (on the bottom) around filaments. Particles within twice $R_{200}$ of \halos\ more massive than $10^{11} \hMsun$ have been excised and the radial flow into clusters has been subtracted (see text for details).}
\label{fig:vel_field}
\end{figure}
This difference between gas and dark matter is more pronounced in profiles of gas fractions ($\rm{fraction} = \frac{\rho_{\rm{gas}}}{\rho_{\rm{gas}} + \rho_{\rm{DM}}}$), which we show in the bottom panel of Fig. \ref{fig:overdensity}: while the gas fraction rises towards filaments as expected, a notable {\it bump} at a distance of around 2-3 Mpc from the filament is evidently appears. Its peak fraction varies with the distance to the node, indicated by the coloured lines. This increased gas fraction closer to the cluster (the "bump") seems to be related to a turbulent interaction generated by the convergence of gas falling into the cluster through filaments and gas falling in isotropically. In a later section (Sec. \ref{sec:subsection4.1}) we argue that this feature originates from the different dynamical signatures of gas and dark matter, as will become evident through their velocity fields. Without experiencing pressure and contractions from cooling, dark matter is able to reach higher velocities. 
Note, however, that the increase of gas fractions towards filaments seen here is not found in a study of filaments at larger scales and less dense environments by \citet{Gheller2019}. In their study, the authors found that the baryon fraction enclosed in filaments normalised by the cosmic mean depends with the filament mass, barely varies with different feedback models and tends to be below the unity at around $0.9$. One reason for this discrepancy could be that we base our filament positions on gas particles. However, differences also could be due to cluster outskirt physics: the dense environments, turbulent mixing of material and thick, perturbed filaments feeding the clusters marking this an unexplored territory.

\subsection{The combined effect of nodes and distance along filaments}
\label{subsec:combined_effect}

Distance to filaments and distance to node are not independent measures. Gas and dark matter densities drop with increasing distance from the nodes (roughly speaking, with distances along the filament spine). Densities therefore drop with orthogonal distances to the filament axes as well as with distances from the node. In addition, densities -- and therefore filament profiles -- might change with the prominence (mass and size) of the node (e.g., groups and clusters). 
In order to investigate this inter-dependence, we split the filament sample into three roughly equal groups according to the size of the node they are connected to. This is comparable to splitting by halo  mass. The three ranges of $R_{200}$ were $(0.0, 0.45) \hMpc$ for the smallest, $(0.45, 0.61) \hMpc$ for medium sized, and $(0.61, 2.25) \hMpc$ for the largest nodes. The last bin includes all central haloes, i.e., the massive cluster at the centre of the simulated regions. Density profiles and gas fractions for different node sizes are shown in the first three columns of Fig. \ref{fig:density_splitbysize} and lines and distance to node bins are the same as in Fig. \ref{fig:overdensity}. As we go to larger-sized nodes, filaments vary more at different radii (distances to the nodes), the typical standard deviation between the profiles up to $d_{\rm fil} = 1 ~ \rm {Mpc} ~ h^{-1}$ is $\approx 0.4$, $\approx 0.5$ and  $\approx 0.63$ for the gas for {\it Small}, {\it Medium}, {\it Large} nodes and $\approx 1.67$, $\approx 2.13$, $\approx 2.56$ for the dark matter case respectively. This means that filaments around clusters are thicker closer to the cluster whereas filaments around smaller nodes show very little variation of thickness. 

As mentioned above, the $d_{\rm node}$ and the distance to the centre of the cluster are not independent of one another. Our goal is to disentangle the contribution of the density profile from the massive node at the centre from the density profile from filaments.  For this, we rotate the true filament networks using random angles and re-calculate the density profiles of these random filaments. Results are shown in the final column of Fig. \ref{fig:density_splitbysize}, labelled "random filaments". The profiles are flat closer to these random filaments and drop off at about $\sim 3\hMpc$ away from them. This suggests that up to $d_{\rm fil} \sim 3\hMpc$, gas is influenced by the filaments themselves. At larger distances from the filaments, density profiles are driven by the clusters (central nodes). The distance to the filament, however, plays a far more important role than the distance to the cluster, which only features as a minor component in the density profiles.

The lower panel in Fig. \ref{fig:density_splitbysize} shows gas fractions as before and reveals that the "bump" (increased gas fractions closer to nodes) at around $2-3 \hMpc$ away from the filament spine we described in the previous section is most prominent close to large, cluster-sized nodes. This is the region around clusters where filament contribution and cluster contribution compete. Further away from the cluster, the "bump" is suppressed. Around smaller nodes (e.g., groups or larger galaxies), this "bump" is much more restrained and constant at all distances from the node. We will explore this feature in more detail in the next section.

\section{The velocity web around galaxy clusters}
\label{Sec:vel_web}
The complex cold gas dynamics of the large scale structure is governed by the successive formation of walls, filaments and clusters \citep{Zeldovich1970, Bond1996, Pichon2011}. In this framework, matter is falling towards a filament from the wall in which the structure is embedded. As a result of these large-scale flows, vortices can form in a specific area perpendicular to the filament axis and in the direction of the wall. Simultaneously, matter flows away from saddle points and towards nodes. This is the complex environment in which galaxies form and evolve. As a consequence, galaxy properties such as orientation of the spin and disk alignments are affected by the winding of the walls into filaments in particular ways \citep{Aragon_Calvo2007, Sousbie2009, Libeskind2012, Codis2012}. 
 
In our study of the velocity field around filaments in \textsc{The ThreeHundred}, we first account for the presence of haloes and other virialized structures (see discussion in Sec. \ref{subsec:haloes_filaments} for the removal process). The removal of particles related to haloes allows us to investigate the motion of the diffuse gas and dark matter towards filaments and -- tangential to it -- towards nodes. The filaments we study converge onto clusters, which leads to a radial pattern of the network. As a consequence, material that is moving towards filaments will simultaneously also move radially towards the cluster. In order to isolate velocities, we first need to control for this systematic bias.

\subsection{Accretion onto clusters: the contribution of filaments}
\label{sec:subsection4.1}

\renewcommand{\thesubfigure}{\roman{subfigure}}
\begin{figure*}
  \centering
  \subfloat[Tangential component of velocity along the filamentary axis separated by different filaments lengths]{\includegraphics[width=0.9\textwidth]{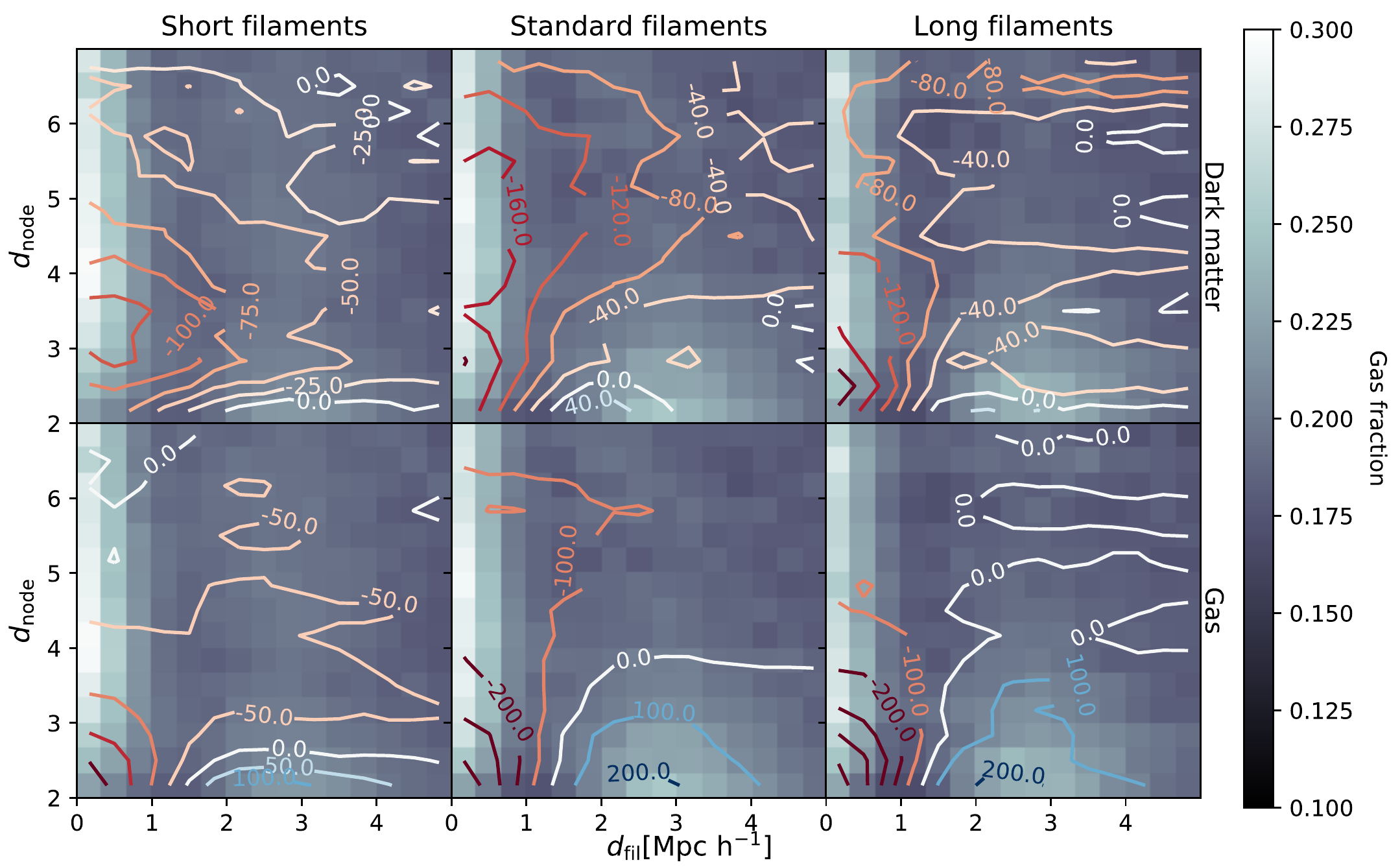}\label{fig:velocity_field1}}
  \vfill
  \subfloat[Tangential component of velocity along the filamentary axis separated by different node sizes]{\includegraphics[width=0.9\textwidth]{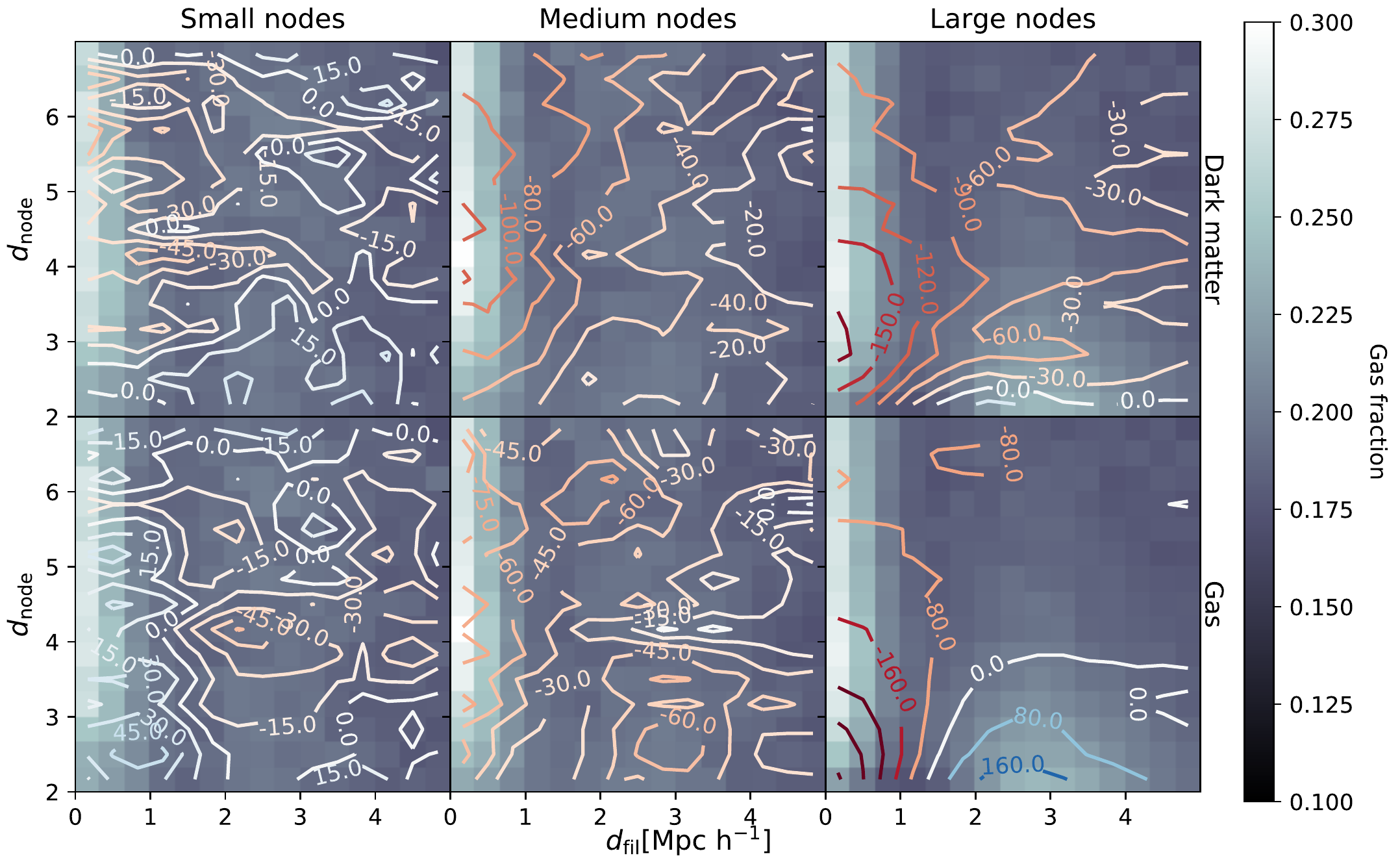}\label{fig:velocity_field2}}
  \caption{Dynamics of gas and dark matter near the centre of the cluster. Top: Tangential component of velocity along the filamentary axis separated by different filaments lengths: short, standard, long. X axis indicates distance to filament in $\rm{h^{-1}Mpc}$, Y axis indicates $d_{\rm{node}}$
Short: Lengths between $2 - 5 ~ \rm{h^{-1}Mpc}$;
standard: $5 - 8 ~ \rm{h^{-1}Mpc}$;
long: $8 - 14 ~ \rm{h^{-1}Mpc}$. Bottom: Tangential component of velocity along the filamentary axis separated by different node sizes: small, medium, big. X axis indicates distance to filament in $\rm{h^{-1}Mpc}$, Y axis indicates $d_{\rm{node}}$.
Small: Node's $R_{200}$ is between $0 - 0.45 ~ \rm{h^{-1}Mpc}$;
medium:  $0.45 - 0.61 ~ \rm{h^{-1}Mpc}$;
large:  $0.61 - 2.3 ~ \rm{h^{-1}Mpc}$.  }
\end{figure*}

\begin{figure}
\centering
\includegraphics[width=\columnwidth]{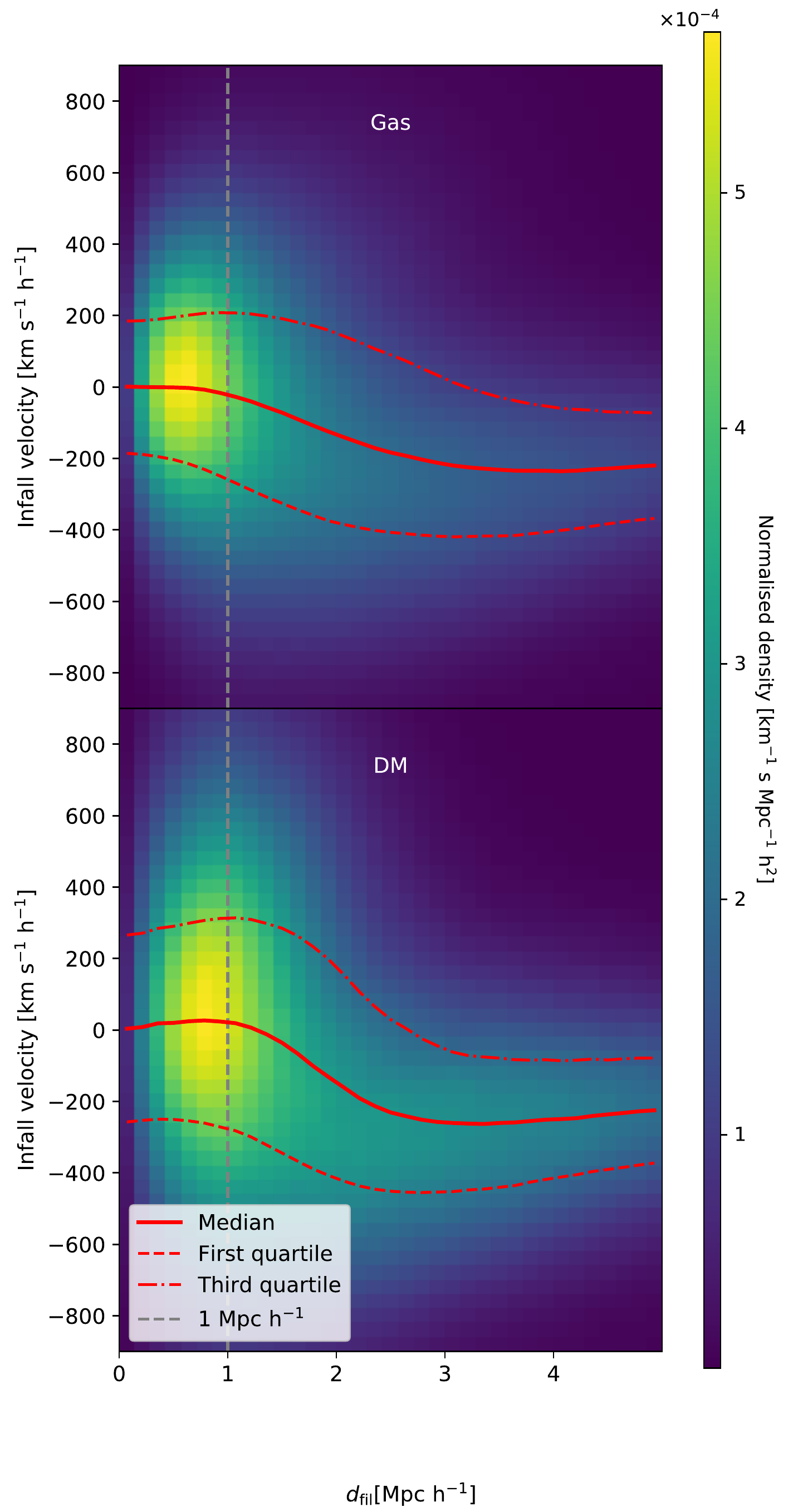}
\caption{Accretion onto filaments: Phase position-velocity diagram for dark matter and gas particles around filaments after removing cluster infall. The x-axis is the distance to the filament axis and the y-axis is the velocity radial component. The red lines indicate the median and the first and third quartiles of the distribution. The vertical grey dashed line indicates a filament thickness of 1 \hMpc. The colour represents the value of the distribution function.}
\label{fig:infall2}
\end{figure}

In order to disentangle the contribution of the filaments from that of the cluster, we first remove the radial component of the infall to the cluster with the following procedure: we transform the velocities to the galaxy cluster's frame of reference. Assuming that the infall velocity depends on the distance to the cluster, we determine the average infall profile in $50$ bins in the range $[0, 10] \hMpc$. This radial velocity field is then used to adjust the velocity of each particle according to its distance from the cluster centre. We do this for gas and dark matter separately.

After removing the cluster's imprint on the global velocity field, we identify the nearest filament segment to each particle and calculate the radial velocity component (defined by the vector from the closest segment to the particle), and tangential velocity component (defined by the direction of the closest filament segment axis, towards the node). As before, we only include complete filaments with both a node and a saddle point and without bifurcation or exit points in this analysis. 

Fig. \ref{fig:vel_field} shows the stacked result of the velocity field of dark matter (top) and gas (bottom) in all $324$ cluster volumes of the sample. It shows the radial and tangential velocity components in bins of distance along the filament spine and in bins of distance to the filament axis. The colour coding indicates the velocity in km/s/h.
Immediately, we can see a smooth collective motion of matter from low density regions towards filaments and nodes. Furthest away from filaments and nodes, the velocity fields of dark matter and gas look similar. However, near the nodes the motion of gas deviates from that of the dark matter: close to the intersection of filaments and nodes, a vortex of gas forms -- which is absent in collisionless dark matter. 
This is consistent with the observation that gas resists the shock in filaments. 

We further investigate this feature in Fig. \ref{fig:velocity_field1} and \ref{fig:velocity_field2}), where we fold in information of gas fractions (shown by the colour-coding) and plot isocontours of the tangential component of the velocity (i.e., towards the nodes and parallel to the filament).
Fig. \ref{fig:velocity_field1} details the velocity flows of gas and dark matter for different filament lengths and \ref{fig:velocity_field2}) shows different node sizes separately.  In this way, we link gas motions to the increase of gas fractions that we first saw in the radial profiles in Fig. \ref{fig:overdensity} and Fig. \ref{fig:density_splitbysize}.
The negative tangential velocity component close to nodes and near filaments in all plots exposes the parallel to the filament motion towards nodes. However, only in the plots of the gas we see a region of positive velocities close to nodes and filaments that coincides with higher gas fractions. This is consistent with our findings in Fig. \ref{fig:overdensity} and \ref{fig:vel_field} and is absent in dark matter. 
The positive velocity region (and hence the gas fraction bump) is seen for all lengths, and it is stronger for the longest filaments bin (Fig. \ref{fig:velocity_field1}) 
and for the large nodes (Fig. \ref{fig:velocity_field2}). Given the dependence of filament lengths and node size (Fig. \ref{fig:hist_length}), this underlines the observation that this phenomenon is related to the central nodes.

\subsection{Accretion onto filaments}
In the previous section we detailed the velocity field in cluster outskirts that is driven by both an infall towards the nodes \textit{and} a collapse of filaments.
In order to isolate the motion towards filaments, we now consider the orthogonal distance to the filament axis and the radial velocity component of gas and dark matter particles. This is shown in Fig. \ref{fig:infall2}, where we plot phase-space diagrams for gas on the left and dark matter on the right hand side.
Both show a similar behaviour: a collective movement of particles falling towards the filament at $\approx 250 h^{-1}$ km/s (indicated by negative velocities) from larger distances, and large random motions inside the filament (as indicated by the vertical dashed lines). 
This plot is quite similar to the ones done stacking haloes  \citep{Hamabata2019, Arthur2019, Oman2013}, but in this case we integrate along the filament axis and stack afterwards.
In this way we see how this structures that as a whole are not yet virialized, when we integrate their properties in a symmetric direction, we recover virial properties. However, this apparent virialized region is more likely to be linked to the motion of material that is shell crossing for the first time, or related to material formerly bounded to haloes.
For dark matter we see a broader but still symmetric range of velocities at the core compared to the gas, but a slight increase in dispersion near the centre. Fig. \ref{fig:overdensity} already illustrated the "fluffier" character of dark matter filaments compared to gas filaments. This is consistent with the ability of dark matter to cross-shell material when it is being accreted to the spine while the gas shocks and cools down. 

Taking this further we split our filament sample into the same three groups as used before, depending on node size. 
Although not shown in this paper, as we move to progressively larger nodes, we observed an increase in the width of the velocity range, as expected for the mass enclosed in those filaments. 

We tested our procedure by repeating the analysis after first randomly rotating the filamentary networks about each central node.  As expected, this completely erases the correlation with the filament's distance, and the radial infall onto the filaments vanishes. 
\begin{figure}
\centering
\includegraphics[width=\columnwidth]{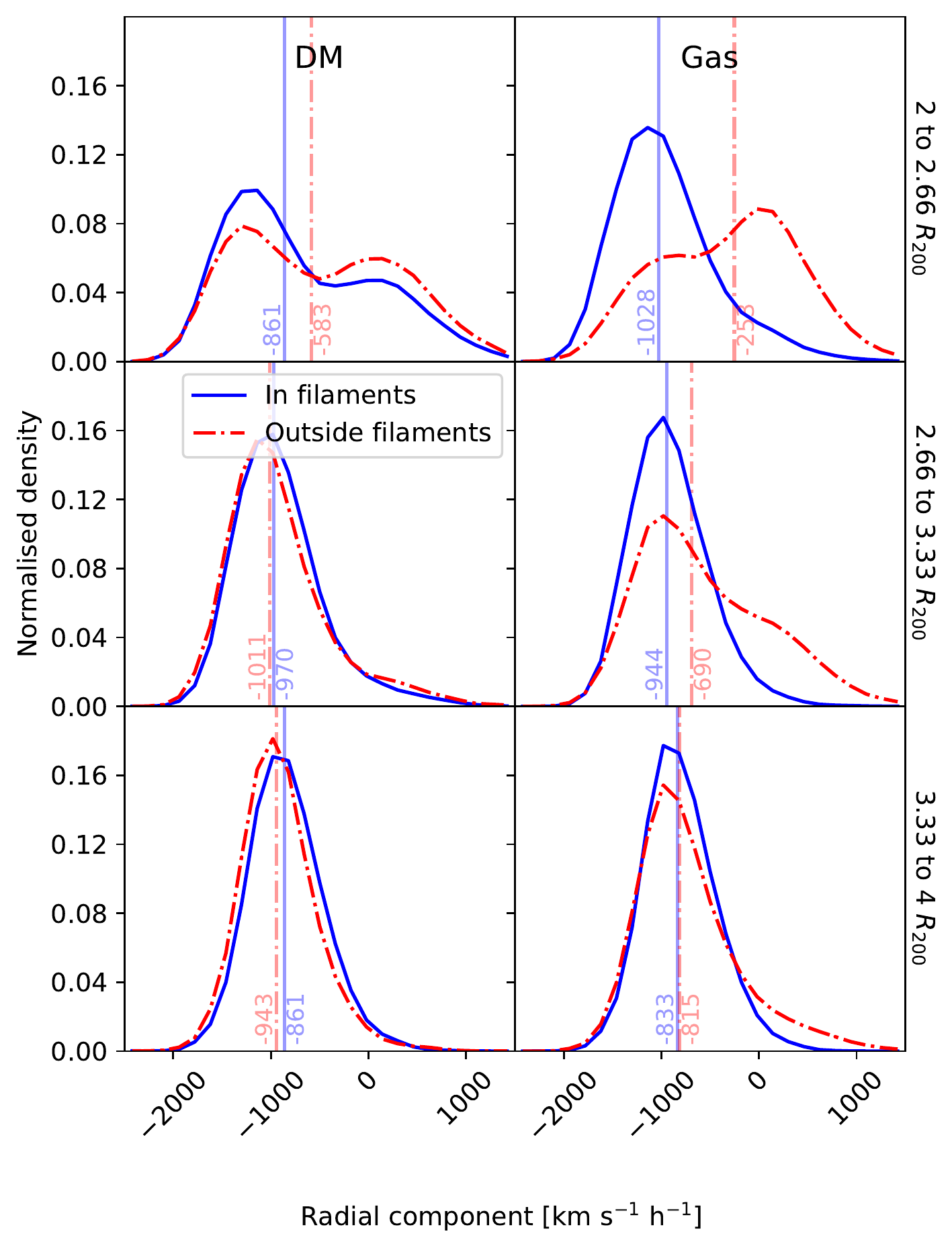}
\caption{Filaments as gas highways: Normalised histogram of the radial infall velocity to the cluster. From top to bottom: particles in the ranges $[2, 2.66]$, $[2.66, 3.33], [3.33, 4] \times R_{200}$ from the centre of the  cluster. On the left: dark matter particles. On the right: gas particles. The lines show particles inside filaments (blue) and outside filaments (red). Their respective median of the distribution is marked with a vertical line of the same colour.
Only relaxed clusters were stacked (relaxedness $ > 1$, see text for description.)}
\label{fig:infall}
\end{figure}

\subsection{Filaments as gas highways}

Finally, we investigate how dark matter and gas move inside and outside filaments as a function of distance to the cluster centre. For this, we only consider galaxy clusters that are dynamically relaxed \citep[as defined by][]{Haggar2020}, since we consider that the an-isotropic contribution of a perturbed halo  will affect our comparison of an isotropic smooth accretion and the one through filaments.
The relaxation parameter is calculated using a combination of the virial ratio, the centre-of-mass offset and the fraction of cluster mass in subhaloes \citep[see also][for details]{Cui2018}, in this section we include clusters with a relaxation parameter greater than one.

In Fig.~\ref{fig:infall} we separate dark matter and gas particles each into two groups: particles {\it inside} filaments have distances to skeletons smaller than $1 \hMpc$ (blue solid line), and particles with larger distances are defined as {\it outside} filaments (red dot-dashed line). To obtain the curves, we calculated radial velocities for all particles and stacked results of all relaxed clusters. From top to bottom, we show radial velocity components in three shells of increasing distance from the cluster centre. The inner and outer radii of these shells are $[2, 2.66)$, $[2.66, 3.33)$, and $[3.33, 4)$. As before, all particles within $2xR_{200}$ of haloes are excluded from this analysis. 

Furthest away from the cluster (bottom left panel in Fig~\ref{fig:infall}), dark matter and gas move towards nodes at a rate that is largely independent of whether or not a particle is inside or outside a filament, the medians of the distributions, marked as vertical lines of the same colours, indicate velocities around $900 h^{-1} km~s^{-1}$. Here, only a small fraction of particles (and more so in gas than in dark matter) has velocities close to $0 h^{-1} km~s^{-1}$ or positive velocities, indicating outflows. 
This changes as we go closer to the cluster: the middle right panel reveals that even at larger distances from the cluster, gas is partially stationary (i.e., this component does not move with respect to the cluster), something not seen in dark matter. Just outside $2 \times R_{200}$ of the node, a clear dichotomy emerges. Both dark matter and gas are split between particles that are moving into the cluster and particles that have a symmetric velocity distribution centred on 0${km ~ s^{-1}}$, which could correspond to an accretion shock just outside the cluster. 
Furthermore, we see increasing differences between gas and dark matter (differences of the median of $\approx 780 h^{-1}\rm{km ~ s^{-1}}$ for the gas, meanwhile for the dark matter it is $\approx 280 h^{-1}\rm{km ~ s^{-1}}$). Gas is falling into nodes with a clear preference to be inside filaments (negative velocities, blue curve), but leaves the cluster with a clear preference of being outside filaments (positive velocities, red curve). For dark matter, this preference is far less obvious.
This observation is consistent with Fig.s \ref{fig:velocity_field1} and \ref{fig:velocity_field2} where we saw a region of negative tangential velocities near the filamentary axis and a region of positive velocities outside the filament. 

The behaviour we see in Fig. \ref{fig:infall}, could be due to random motions of particles in clusters on the one hand, but also to an increasing presence of accretion shock and/or backsplash material close to clusters \citep[e.g.,][]{Haggar2020}. The latter refers to material that has gone through the cluster and will return on a second infall. 
This plot confirms that filaments act as 
highways to funnel material into clusters \citep{Cautun, Kraljic2018}, however with noteable differences between dark matter and gas. 
Gas seems to penetrate the expanding shock shell of clusters more efficiently when it is flowing through a filament, while {\it diffuse} dark matter falling into clusters through filaments does not differ from an isotropic infall. Gas in filaments seems to be less sensitive to shocks, a phenomenon that once motivated the study of cold flows on smaller halo  scales, but also theoretical studies up to cluster scales \citep{Cornuault2018}. Moreover, the different behaviour of gas and dark matter is most prominent close to clusters, 
and becomes less evident further away.

\section{Conclusions}
\label{Sec:conclusions}
In this paper, we have analysed filaments detected by \disperse\ in the environments around a set of 324 cluster re-simulations of \threehundred project. We determined the key characteristics of these structures such as the gas and dark matter profiles and velocity flows.
We clean the filaments excluding all material inside $2 \times R_{200}$ radius of all haloes with masses above $10^{11}$\hMsun. In this way we aim to study unbound material (either gas or dark matter) that constitutes the diffuse material of filaments.
We analyse the filaments by stacking and averaging to describe global properties common to them.
We found evidence that supports the idea that filaments are structures with well defined properties in both dark matter and baryons. Importantly, we identified differences between dark matter and gas velocities close to clusters and just outside filaments. 

The main conclusions of our analysis are summarized in the following:
\begin{itemize}
    \item The dark matter and gas density profiles of filaments surrounding clusters have similar shapes and features, such as the presence of a "core" at the filament axis, a density decay consistent with a power-law in the outskirts and a decay of the peak density moving away from the filament node. We report a characteristic filament thickness of around $0.7 \hMpc$.
    \item The differences between dark matter and gas density profiles become evident in gas fractions: the fraction varies and increases towards the filament spine. Furthermore, a local increase at roughly $2 \hMpc$ distance from the cluster in the gas fraction (a "bump") suggests turbulent gas that is not present in dark matter.
    \item Filaments are surrounded by a structured velocity field largely as predicted by theory. This needs to be carefully disentangled from the general velocity inflow surrounding large clusters.
    \item The velocity-space phase diagram of filaments is analogous to that of the clusters. The matterial infalls at around $-250 h^{-1}$km/s in the outskirts and tends to be centred at zero at the spine. There are clear differences between the motion of gas and dark matter particles. The velocity dispersion is present in the whole range, but it is higher for the dark matter, showing an increment towards the spine. Meanwhile, due to gas physics, the velocity dispersion of the gas decreases towards the filament.
    \item Filaments form highways along which material preferentially flows into the clusters. We find that the material that is conducted through the filament has higher infall velocity towards the cluster centre than the material falling in other directions. This is particularly noticeable for the gas, most likely due to the gas pressure, in line with the idea of filaments as channels that allow the infall of fresh material towards the central region.
\end{itemize}

While higher resolution simulations will be able to better constrain our findings, 
the results clearly point to different paths of dark matter and gas flows towards the central region of galaxy clusters.

\section*{Acknowledgements}
This work has been made possible by \threehundred collaboration\footnote{https://www.the300-project.org}, which benefits from financial support of the European Union’s Horizon 2020 Research and Innovation programme under the Marie Sk\l{}odowskaw-Curie grant agreement number 734374, i.e. the LACEGAL project. \threehundred simulations used in this paper have been performed in the MareNostrum Supercomputer at the Barcelona Supercomputing Center, thanks to CPU time granted by the Red Espa\~{n}ola de Supercomputaci\'{o}n.
UK acknowledges support from STFC. 
GY and AK acknowledge financial support from \textit{Ministerio de
Ciencia, Innovación y Universidades / Fondo Europeo de Desarrollo
Regional}, under research grant PGC2018-094975-C21.
WC acknowledges support from the European Research Council under grant number 670193 (the COSFORM project). FS and AR acknowledge the financial support from Consejo  Nacional de Investigaciones Científicas y Técnicas (CONICET,  Argentina) by PICT-2016-4174 grant and Secretaría de Ciencia y Tecnología de la Universidad Nacional de Córdoba (SeCyT-UNC, Argentina) by Consolidar-2018-2020 grant.

The authors of this paper contributed in the following ways: AR, FS, FRP, UK \& MEG formed the core team. AR performed the analysis which was largely completed during an EU (LACEGAL) exchange visit to the University of Nottingham. CW provided insightful comments that added to the interpretation of the results and ran \disperse. GY undertook the simulations with AK generating the halo  catalogues.

\section*{Data Availability}
Data available on request to \threehundred collaboration\footnote{\url{www.the300-project.org}}.





\bibliographystyle{mnras}
\bibliography{biblio}

\begin{thebibliography}{}
\makeatletter
\relax
\def\mn@urlcharsother{\let\do\@makeother \do\$\do\&\do\#\do\^\do\_\do\%\do\~}
\def\mn@doi{\begingroup\mn@urlcharsother \@ifnextchar [ {\mn@doi@}
  {\mn@doi@[]}}
\def\mn@doi@[#1]#2{\def\@tempa{#1}\ifx\@tempa\@empty \href
  {http://dx.doi.org/#2} {doi:#2}\else \href {http://dx.doi.org/#2} {#1}\fi
  \endgroup}
\def\mn@eprint#1#2{\mn@eprint@#1:#2::\@nil}
\def\mn@eprint@arXiv#1{\href {http://arxiv.org/abs/#1} {{\tt arXiv:#1}}}
\def\mn@eprint@dblp#1{\href {http://dblp.uni-trier.de/rec/bibtex/#1.xml}
  {dblp:#1}}
\def\mn@eprint@#1:#2:#3:#4\@nil{\def\@tempa {#1}\def\@tempb {#2}\def\@tempc
  {#3}\ifx \@tempc \@empty \let \@tempc \@tempb \let \@tempb \@tempa \fi \ifx
  \@tempb \@empty \def\@tempb {arXiv}\fi \@ifundefined
  {mn@eprint@\@tempb}{\@tempb:\@tempc}{\expandafter \expandafter \csname
  mn@eprint@\@tempb\endcsname \expandafter{\@tempc}}}

\bibitem[\protect\citeauthoryear{{Arag{\'o}n-Calvo}, {Jones}, {van de Weygaert}
   \& {van der Hulst}}{{Arag{\'o}n-Calvo} et~al.}{2007}]{Aragon_Calvo2007}
{Arag{\'o}n-Calvo} M.~A.,  {Jones} B.~J.~T.,  {van de Weygaert} R.,   {van der
  Hulst} J.~M.,  2007, \mn@doi [\aap] {10.1051/0004-6361:20077880}, \href
  {https://ui.adsabs.harvard.edu/abs/2007A&A...474..315A} {474, 315}

\bibitem[\protect\citeauthoryear{{Arag{\'o}n-Calvo}, {van de Weygaert}  \&
  {Jones}}{{Arag{\'o}n-Calvo} et~al.}{2010}]{Aragon_Calvo2010}
{Arag{\'o}n-Calvo} M.~A.,  {van de Weygaert} R.,   {Jones} B. J.~T.,  2010,
  \mn@doi [\mnras] {10.1111/j.1365-2966.2010.17263.x}, \href
  {https://ui.adsabs.harvard.edu/abs/2010MNRAS.408.2163A} {408, 2163}

\bibitem[\protect\citeauthoryear{{Arnold}, {Shandarin}  \&
  {Zeldovich}}{{Arnold} et~al.}{1982}]{Arnold1982}
{Arnold} V.~I.,  {Shandarin} S.~F.,   {Zeldovich} I.~B.,  1982, \mn@doi
  [Geophysical and Astrophysical Fluid Dynamics] {10.1080/03091928208209001},
  \href {https://ui.adsabs.harvard.edu/abs/1982GApFD..20..111A} {20, 111}

\bibitem[\protect\citeauthoryear{{Arthur} et~al.,}{{Arthur}
  et~al.}{2019}]{Arthur2019}
{Arthur} J.,  et~al., 2019, \mn@doi [\mnras] {10.1093/mnras/stz212}, \href
  {https://ui.adsabs.harvard.edu/abs/2019MNRAS.484.3968A} {484, 3968}

\bibitem[\protect\citeauthoryear{{Beck} et~al.,}{{Beck}
  et~al.}{2016}]{Beck2016}
{Beck} A.~M.,  et~al., 2016, \mn@doi [\mnras] {10.1093/mnras/stv2443}, \href
  {https://ui.adsabs.harvard.edu/abs/2016MNRAS.455.2110B} {455, 2110}

\bibitem[\protect\citeauthoryear{{Behroozi}, {Wechsler}  \& {Wu}}{{Behroozi}
  et~al.}{2013}]{rockstar}
{Behroozi} P.~S.,  {Wechsler} R.~H.,   {Wu} H.-Y.,  2013, \mn@doi [\apj]
  {10.1088/0004-637X/762/2/109}, \href
  {https://ui.adsabs.harvard.edu/abs/2013ApJ...762..109B} {762, 109}

\bibitem[\protect\citeauthoryear{{Benitez-Llambay}}{{Benitez-Llambay}}{2015}]{Benitez_Llambay2015}
{Benitez-Llambay} A.,  2015, {Py-Sphviewer: Py-Sphviewer V1.0.0},
  \mn@doi{10.5281/zenodo.21703}

\bibitem[\protect\citeauthoryear{{Birnboim} \& {Dekel}}{{Birnboim} \&
  {Dekel}}{2003}]{Birnboim2003}
{Birnboim} Y.,  {Dekel} A.,  2003, \mn@doi [\mnras]
  {10.1046/j.1365-8711.2003.06955.x}, \href
  {https://ui.adsabs.harvard.edu/abs/2003MNRAS.345..349B} {345, 349}

\bibitem[\protect\citeauthoryear{{Bonamente}, {Nevalainen}, {Tilton},
  {Liivam{\"a}gi}, {Tempel}, {Hein{\"a}m{\"a}ki}  \& {Fang}}{{Bonamente}
  et~al.}{2016}]{Bonamente}
{Bonamente} M.,  {Nevalainen} J.,  {Tilton} E.,  {Liivam{\"a}gi} J.,  {Tempel}
  E.,  {Hein{\"a}m{\"a}ki} P.,   {Fang} T.,  2016, \mn@doi [\mnras]
  {10.1093/mnras/stw285}, \href
  {https://ui.adsabs.harvard.edu/abs/2016MNRAS.457.4236B} {457, 4236}

\bibitem[\protect\citeauthoryear{{Bond}, {Kofman}  \& {Pogosyan}}{{Bond}
  et~al.}{1996}]{Bond1996}
{Bond} J.~R.,  {Kofman} L.,   {Pogosyan} D.,  1996, \mn@doi [\nat]
  {10.1038/380603a0}, \href
  {https://ui.adsabs.harvard.edu/abs/1996Natur.380..603B} {380, 603}

\bibitem[\protect\citeauthoryear{{Bond}, {Strauss}  \& {Cen}}{{Bond}
  et~al.}{2010}]{Bond2010}
{Bond} N.~A.,  {Strauss} M.~A.,   {Cen} R.,  2010, \mn@doi [\mnras]
  {10.1111/j.1365-2966.2010.17307.x}, \href
  {https://ui.adsabs.harvard.edu/abs/2010MNRAS.409..156B} {409, 156}

\bibitem[\protect\citeauthoryear{{Bonjean}, {Aghanim}, {Douspis}, {Malavasi}
  \& {Tanimura}}{{Bonjean} et~al.}{2019}]{Bonjean2019}
{Bonjean} V.,  {Aghanim} N.,  {Douspis} M.,  {Malavasi} N.,   {Tanimura} H.,
  2019, arXiv e-prints, \href
  {https://ui.adsabs.harvard.edu/abs/2019arXiv191206559B} {p. arXiv:1912.06559}

\bibitem[\protect\citeauthoryear{{Bykov}, {Paerels}  \& {Petrosian}}{{Bykov}
  et~al.}{2008}]{Bykov2008}
{Bykov} A.~M.,  {Paerels} F.~B.~S.,   {Petrosian} V.,  2008, \mn@doi [\ssr]
  {10.1007/s11214-008-9309-4}, \href
  {https://ui.adsabs.harvard.edu/abs/2008SSRv..134..141B} {134, 141}

\bibitem[\protect\citeauthoryear{{Cautun}, {van de Weygaert}  \&
  {Jones}}{{Cautun} et~al.}{2013}]{nexus1}
{Cautun} M.,  {van de Weygaert} R.,   {Jones} B. J.~T.,  2013, \mn@doi [\mnras]
  {10.1093/mnras/sts416}, \href
  {https://ui.adsabs.harvard.edu/abs/2013MNRAS.429.1286C} {429, 1286}

\bibitem[\protect\citeauthoryear{{Cautun}, {van de Weygaert}, {Jones}  \&
  {Frenk}}{{Cautun} et~al.}{2014}]{Cautun}
{Cautun} M.,  {van de Weygaert} R.,  {Jones} B. J.~T.,   {Frenk} C.~S.,  2014,
  \mn@doi [\mnras] {10.1093/mnras/stu768}, \href
  {https://ui.adsabs.harvard.edu/abs/2014MNRAS.441.2923C} {441, 2923}

\bibitem[\protect\citeauthoryear{{Cen} \& {Ostriker}}{{Cen} \&
  {Ostriker}}{1999}]{Ostriker}
{Cen} R.,  {Ostriker} J.~P.,  1999, \mn@doi [\apj] {10.1086/306949}, \href
  {https://ui.adsabs.harvard.edu/abs/1999ApJ...514....1C} {514, 1}

\bibitem[\protect\citeauthoryear{{Codis}, {Pichon}, {Devriendt}, {Slyz},
  {Pogosyan}, {Dubois}  \& {Sousbie}}{{Codis} et~al.}{2012}]{Codis2012}
{Codis} S.,  {Pichon} C.,  {Devriendt} J.,  {Slyz} A.,  {Pogosyan} D.,
  {Dubois} Y.,   {Sousbie} T.,  2012, \mn@doi [\mnras]
  {10.1111/j.1365-2966.2012.21636.x}, \href
  {https://ui.adsabs.harvard.edu/abs/2012MNRAS.427.3320C} {427, 3320}

\bibitem[\protect\citeauthoryear{{Codis}, {Pichon}  \& {Pogosyan}}{{Codis}
  et~al.}{2015}]{Codis2015}
{Codis} S.,  {Pichon} C.,   {Pogosyan} D.,  2015, \mn@doi [\mnras]
  {10.1093/mnras/stv1570}, \href
  {https://ui.adsabs.harvard.edu/abs/2015MNRAS.452.3369C} {452, 3369}

\bibitem[\protect\citeauthoryear{{Colberg}, {Krughoff}  \&
  {Connolly}}{{Colberg} et~al.}{2005}]{Colberg}
{Colberg} J.~M.,  {Krughoff} K.~S.,   {Connolly} A.~J.,  2005, \mn@doi [\mnras]
  {10.1111/j.1365-2966.2005.08897.x}, \href
  {https://ui.adsabs.harvard.edu/abs/2005MNRAS.359..272C} {359, 272}

\bibitem[\protect\citeauthoryear{{Cornuault}, {Lehnert}, {Boulanger}  \&
  {Guillard}}{{Cornuault} et~al.}{2018}]{Cornuault2018}
{Cornuault} N.,  {Lehnert} M.~D.,  {Boulanger} F.,   {Guillard} P.,  2018,
  \mn@doi [\aap] {10.1051/0004-6361/201629229}, \href
  {https://ui.adsabs.harvard.edu/abs/2018A&A...610A..75C} {610, A75}

\bibitem[\protect\citeauthoryear{{Cui}, {Knebe}, {Yepes}, {Yang}, {Borgani},
  {Kang}, {Power}  \& {Staveley-Smith}}{{Cui} et~al.}{2018a}]{Cui2018b}
{Cui} W.,  {Knebe} A.,  {Yepes} G.,  {Yang} X.,  {Borgani} S.,  {Kang} X.,
  {Power} C.,   {Staveley-Smith} L.,  2018a, \mn@doi [\mnras]
  {10.1093/mnras/stx2323}, \href
  {https://ui.adsabs.harvard.edu/abs/2018MNRAS.473...68C} {473, 68}

\bibitem[\protect\citeauthoryear{{Cui} et~al.,}{{Cui} et~al.}{2018b}]{Cui2018}
{Cui} W.,  et~al., 2018b, \mn@doi [\mnras] {10.1093/mnras/sty2111}, \href
  {https://ui.adsabs.harvard.edu/abs/2018MNRAS.480.2898C} {480, 2898}

\bibitem[\protect\citeauthoryear{{Cui} et~al.,}{{Cui} et~al.}{2019}]{Cui2019}
{Cui} W.,  et~al., 2019, \mn@doi [\mnras] {10.1093/mnras/stz565}, \href
  {https://ui.adsabs.harvard.edu/abs/2019MNRAS.485.2367C} {485, 2367}

\bibitem[\protect\citeauthoryear{{Danovich}, {Dekel}, {Hahn}  \&
  {Teyssier}}{{Danovich} et~al.}{2012}]{Danovich2012}
{Danovich} M.,  {Dekel} A.,  {Hahn} O.,   {Teyssier} R.,  2012, \mn@doi
  [\mnras] {10.1111/j.1365-2966.2012.20751.x}, \href
  {https://ui.adsabs.harvard.edu/abs/2012MNRAS.422.1732D} {422, 1732}

\bibitem[\protect\citeauthoryear{{Dav{\'e}} et~al.,}{{Dav{\'e}}
  et~al.}{2001}]{Dave}
{Dav{\'e}} R.,  et~al., 2001, \mn@doi [\apj] {10.1086/320548}, \href
  {https://ui.adsabs.harvard.edu/abs/2001ApJ...552..473D} {552, 473}

\bibitem[\protect\citeauthoryear{{Dekel} et~al.,}{{Dekel}
  et~al.}{2009}]{Dekel2009}
{Dekel} A.,  et~al., 2009, \mn@doi [\nat] {10.1038/nature07648}, \href
  {https://ui.adsabs.harvard.edu/abs/2009Natur.457..451D} {457, 451}

\bibitem[\protect\citeauthoryear{{Dolag}, {Meneghetti}, {Moscardini}, {Rasia}
  \& {Bonaldi}}{{Dolag} et~al.}{2006}]{Dolag}
{Dolag} K.,  {Meneghetti} M.,  {Moscardini} L.,  {Rasia} E.,   {Bonaldi} A.,
  2006, \mn@doi [\mnras] {10.1111/j.1365-2966.2006.10511.x}, \href
  {https://ui.adsabs.harvard.edu/abs/2006MNRAS.370..656D} {370, 656}

\bibitem[\protect\citeauthoryear{{Dubois}, {Pichon}, {Haehnelt}, {Kimm},
  {Slyz}, {Devriendt}  \& {Pogosyan}}{{Dubois} et~al.}{2012}]{Dubois2012}
{Dubois} Y.,  {Pichon} C.,  {Haehnelt} M.,  {Kimm} T.,  {Slyz} A.,  {Devriendt}
  J.,   {Pogosyan} D.,  2012, \mn@doi [\mnras]
  {10.1111/j.1365-2966.2012.21160.x}, \href
  {https://ui.adsabs.harvard.edu/abs/2012MNRAS.423.3616D} {423, 3616}

\bibitem[\protect\citeauthoryear{{Eckert} et~al.,}{{Eckert}
  et~al.}{2015}]{Eckert}
{Eckert} D.,  et~al., 2015, \mn@doi [\nat] {10.1038/nature16058}, \href
  {https://ui.adsabs.harvard.edu/abs/2015Natur.528..105E} {528, 105}

\bibitem[\protect\citeauthoryear{{Forero-Romero}, {Hoffman}, {Gottl{\"o}ber},
  {Klypin}  \& {Yepes}}{{Forero-Romero} et~al.}{2009}]{Forero_Romero2009}
{Forero-Romero} J.~E.,  {Hoffman} Y.,  {Gottl{\"o}ber} S.,  {Klypin} A.,
  {Yepes} G.,  2009, \mn@doi [\mnras] {10.1111/j.1365-2966.2009.14885.x}, \href
  {https://ui.adsabs.harvard.edu/abs/2009MNRAS.396.1815F} {396, 1815}

\bibitem[\protect\citeauthoryear{{Gal{\'a}rraga-Espinosa}, {Aghanim}, {Langer},
  {Gouin}  \& {Malavasi}}{{Gal{\'a}rraga-Espinosa}
  et~al.}{2020}]{Galarraga_Espinosa2020}
{Gal{\'a}rraga-Espinosa} D.,  {Aghanim} N.,  {Langer} M.,  {Gouin} C.,
  {Malavasi} N.,  2020, arXiv e-prints, \href
  {https://ui.adsabs.harvard.edu/abs/2020arXiv200309697G} {p. arXiv:2003.09697}

\bibitem[\protect\citeauthoryear{{Ganeshaiah~Veena}, Cautun, Tempel, van~de
  Weygaert  \& Frenk}{{Ganeshaiah~Veena} et~al.}{2019}]{Veena2019}
{Ganeshaiah~Veena} P.,  Cautun M.,  Tempel E.,  van~de Weygaert R.,   Frenk
  C.~S.,  2019, \mn@doi [Monthly Notices of the Royal Astronomical Society]
  {10.1093/mnras/stz1343}, 487, 1607

\bibitem[\protect\citeauthoryear{{Gheller} \& {Vazza}}{{Gheller} \&
  {Vazza}}{2019}]{Gheller2019}
{Gheller} C.,  {Vazza} F.,  2019, \mn@doi [\mnras] {10.1093/mnras/stz843},
  \href {https://ui.adsabs.harvard.edu/abs/2019MNRAS.486..981G} {486, 981}

\bibitem[\protect\citeauthoryear{{Gheller}, {Vazza}, {Favre}  \&
  {Br{\"u}ggen}}{{Gheller} et~al.}{2015}]{Gheller2015}
{Gheller} C.,  {Vazza} F.,  {Favre} J.,   {Br{\"u}ggen} M.,  2015, \mn@doi
  [\mnras] {10.1093/mnras/stv1646}, \href
  {https://ui.adsabs.harvard.edu/abs/2015MNRAS.453.1164G} {453, 1164}

\bibitem[\protect\citeauthoryear{{Gurbatov}, {Saichev}  \&
  {Shandarin}}{{Gurbatov} et~al.}{1989}]{Gurbatov1989}
{Gurbatov} S.~N.,  {Saichev} A.~I.,   {Shandarin} S.~F.,  1989, \mn@doi
  [\mnras] {10.1093/mnras/236.2.385}, \href
  {https://ui.adsabs.harvard.edu/abs/1989MNRAS.236..385G} {236, 385}

\bibitem[\protect\citeauthoryear{{Haggar}, {Gray}, {Pearce}, {Knebe}, {Cui},
  {Mostoghiu}  \& {Yepes}}{{Haggar} et~al.}{2020}]{Haggar2020}
{Haggar} R.,  {Gray} M.~E.,  {Pearce} F.~R.,  {Knebe} A.,  {Cui} W.,
  {Mostoghiu} R.,   {Yepes} G.,  2020, \mn@doi [\mnras]
  {10.1093/mnras/staa273}, \href
  {https://ui.adsabs.harvard.edu/abs/2020MNRAS.492.6074H} {492, 6074}

\bibitem[\protect\citeauthoryear{{Hamabata}, {Oguri}  \&
  {Nishimichi}}{{Hamabata} et~al.}{2019}]{Hamabata2019}
{Hamabata} A.,  {Oguri} M.,   {Nishimichi} T.,  2019, \mn@doi [\mnras]
  {10.1093/mnras/stz2227}, \href
  {https://ui.adsabs.harvard.edu/abs/2019MNRAS.489.1344H} {489, 1344}

\bibitem[\protect\citeauthoryear{{Hidding}, {Shandarin}  \& {van de
  Weygaert}}{{Hidding} et~al.}{2014}]{Hidding2014}
{Hidding} J.,  {Shandarin} S.~F.,   {van de Weygaert} R.,  2014, \mn@doi
  [\mnras] {10.1093/mnras/stt2142}, \href
  {https://ui.adsabs.harvard.edu/abs/2014MNRAS.437.3442H} {437, 3442}

\bibitem[\protect\citeauthoryear{{Klypin}, {Yepes}, {Gottl{\"o}ber}, {Prada}
  \& {He{\ss}}}{{Klypin} et~al.}{2016}]{multidark}
{Klypin} A.,  {Yepes} G.,  {Gottl{\"o}ber} S.,  {Prada} F.,   {He{\ss}} S.,
  2016, \mn@doi [\mnras] {10.1093/mnras/stw248}, \href
  {https://ui.adsabs.harvard.edu/abs/2016MNRAS.457.4340K} {457, 4340}

\bibitem[\protect\citeauthoryear{{Knollmann} \& {Knebe}}{{Knollmann} \&
  {Knebe}}{2009}]{AHF}
{Knollmann} S.~R.,  {Knebe} A.,  2009, \mn@doi [\apjs]
  {10.1088/0067-0049/182/2/608}, \href
  {https://ui.adsabs.harvard.edu/abs/2009ApJS..182..608K} {182, 608}

\bibitem[\protect\citeauthoryear{{Kraljic} et~al.,}{{Kraljic}
  et~al.}{2018}]{Kraljic2018}
{Kraljic} K.,  et~al., 2018, \mn@doi [\mnras] {10.1093/mnras/stx2638}, \href
  {https://ui.adsabs.harvard.edu/abs/2018MNRAS.474..547K} {474, 547}

\bibitem[\protect\citeauthoryear{{Kraljic} et~al.,}{{Kraljic}
  et~al.}{2019}]{Kraljic2019}
{Kraljic} K.,  et~al., 2019, \mn@doi [\mnras] {10.1093/mnras/sty3216}, \href
  {https://ui.adsabs.harvard.edu/abs/2019MNRAS.483.3227K} {483, 3227}

\bibitem[\protect\citeauthoryear{Kuchner et~al.,}{Kuchner
  et~al.}{2020}]{Kuchner2020}
Kuchner U.,  et~al., 2020, \mn@doi [Monthly Notices of the Royal Astronomical
  Society] {10.1093/mnras/staa1083}, 494, 5473

\bibitem[\protect\citeauthoryear{{Laigle} et~al.,}{{Laigle}
  et~al.}{2015}]{Laigle2015}
{Laigle} C.,  et~al., 2015, \mn@doi [\mnras] {10.1093/mnras/stu2289}, \href
  {https://ui.adsabs.harvard.edu/abs/2015MNRAS.446.2744L} {446, 2744}

\bibitem[\protect\citeauthoryear{{Libeskind}, {Hoffman}, {Knebe}, {Steinmetz},
  {Gottl{\"o}ber}, {Metuki}  \& {Yepes}}{{Libeskind}
  et~al.}{2012}]{Libeskind2012}
{Libeskind} N.~I.,  {Hoffman} Y.,  {Knebe} A.,  {Steinmetz} M.,
  {Gottl{\"o}ber} S.,  {Metuki} O.,   {Yepes} G.,  2012, \mn@doi [\mnras]
  {10.1111/j.1745-3933.2012.01222.x}, \href
  {https://ui.adsabs.harvard.edu/abs/2012MNRAS.421L.137L} {421, L137}

\bibitem[\protect\citeauthoryear{{Libeskind}, {Tempel}, {Hoffman}, {Tully}  \&
  {Courtois}}{{Libeskind} et~al.}{2015}]{Libeskind2015}
{Libeskind} N.~I.,  {Tempel} E.,  {Hoffman} Y.,  {Tully} R.~B.,   {Courtois}
  H.,  2015, \mn@doi [\mnras] {10.1093/mnrasl/slv099}, \href
  {https://ui.adsabs.harvard.edu/abs/2015MNRAS.453L.108L} {453, L108}

\bibitem[\protect\citeauthoryear{{Libeskind} et~al.,}{{Libeskind}
  et~al.}{2018}]{Libeskind}
{Libeskind} N.~I.,  et~al., 2018, \mn@doi [\mnras] {10.1093/mnras/stx1976},
  \href {https://ui.adsabs.harvard.edu/abs/2018MNRAS.473.1195L} {473, 1195}

\bibitem[\protect\citeauthoryear{{Malavasi} et~al.,}{{Malavasi}
  et~al.}{2017}]{Malavasi2017}
{Malavasi} N.,  et~al., 2017, \mn@doi [\mnras] {10.1093/mnras/stw2864}, \href
  {https://ui.adsabs.harvard.edu/abs/2017MNRAS.465.3817M} {465, 3817}

\bibitem[\protect\citeauthoryear{{Malavasi}, {Aghanim}, {Douspis}, {Tanimura}
  \& {Bonjean}}{{Malavasi} et~al.}{2020}]{Malavasi2020}
{Malavasi} N.,  {Aghanim} N.,  {Douspis} M.,  {Tanimura} H.,   {Bonjean} V.,
  2020, arXiv e-prints, \href
  {https://ui.adsabs.harvard.edu/abs/2020arXiv200201486M} {p. arXiv:2002.01486}

\bibitem[\protect\citeauthoryear{{Mart{\'\i}nez}, {Muriel}  \&
  {Coenda}}{{Mart{\'\i}nez} et~al.}{2016}]{Martinez}
{Mart{\'\i}nez} H.~J.,  {Muriel} H.,   {Coenda} V.,  2016, \mn@doi [\mnras]
  {10.1093/mnras/stv2295}, \href
  {https://ui.adsabs.harvard.edu/abs/2016MNRAS.455..127M} {455, 127}

\bibitem[\protect\citeauthoryear{{Murante}, {Monaco}, {Giovalli}, {Borgani}  \&
  {Diaferio}}{{Murante} et~al.}{2010}]{Murante2010}
{Murante} G.,  {Monaco} P.,  {Giovalli} M.,  {Borgani} S.,   {Diaferio} A.,
  2010, \mn@doi [\mnras] {10.1111/j.1365-2966.2010.16567.x}, \href
  {https://ui.adsabs.harvard.edu/abs/2010MNRAS.405.1491M} {405, 1491}

\bibitem[\protect\citeauthoryear{{Navarro}, {Frenk}  \& {White}}{{Navarro}
  et~al.}{1996}]{NFW}
{Navarro} J.~F.,  {Frenk} C.~S.,   {White} S. D.~M.,  1996, \mn@doi [\apj]
  {10.1086/177173}, \href
  {https://ui.adsabs.harvard.edu/abs/1996ApJ...462..563N} {462, 563}

\bibitem[\protect\citeauthoryear{{Nicastro} et~al.,}{{Nicastro}
  et~al.}{2013}]{Nicastro}
{Nicastro} F.,  et~al., 2013, \mn@doi [\apj] {10.1088/0004-637X/769/2/90},
  \href {https://ui.adsabs.harvard.edu/abs/2013ApJ...769...90N} {769, 90}

\bibitem[\protect\citeauthoryear{{Oman}, {Hudson}  \& {Behroozi}}{{Oman}
  et~al.}{2013}]{Oman2013}
{Oman} K.~A.,  {Hudson} M.~J.,   {Behroozi} P.~S.,  2013, \mn@doi [\mnras]
  {10.1093/mnras/stt328}, \href
  {https://ui.adsabs.harvard.edu/abs/2013MNRAS.431.2307O} {431, 2307}

\bibitem[\protect\citeauthoryear{{Pereyra}, {Sgr{\'o}}, {Merch{\'a}n},
  {Stasyszyn}  \& {Paz}}{{Pereyra} et~al.}{2019}]{pereyra}
{Pereyra} L.~A.,  {Sgr{\'o}} M.~A.,  {Merch{\'a}n} M.~E.,  {Stasyszyn} F.~A.,
  {Paz} D.~J.,  2019, arXiv e-prints, \href
  {https://ui.adsabs.harvard.edu/abs/2019arXiv191106768P} {p. arXiv:1911.06768}

\bibitem[\protect\citeauthoryear{{Pichon}, {Pogosyan}, {Kimm}, {Slyz},
  {Devriendt}  \& {Dubois}}{{Pichon} et~al.}{2011}]{Pichon2011}
{Pichon} C.,  {Pogosyan} D.,  {Kimm} T.,  {Slyz} A.,  {Devriendt} J.,
  {Dubois} Y.,  2011, \mn@doi [\mnras] {10.1111/j.1365-2966.2011.19640.x},
  \href {https://ui.adsabs.harvard.edu/abs/2011MNRAS.418.2493P} {418, 2493}

\bibitem[\protect\citeauthoryear{{Power} et~al.,}{{Power}
  et~al.}{2020}]{Power2020}
{Power} C.,  et~al., 2020, \mn@doi [\mnras] {10.1093/mnras/stz3176}, \href
  {https://ui.adsabs.harvard.edu/abs/2020MNRAS.491.3923P} {491, 3923}

\bibitem[\protect\citeauthoryear{{Rasia} et~al.,}{{Rasia}
  et~al.}{2015}]{Rasia2015}
{Rasia} E.,  et~al., 2015, \mn@doi [\apjl] {10.1088/2041-8205/813/1/L17}, \href
  {https://ui.adsabs.harvard.edu/abs/2015ApJ...813L..17R} {813, L17}

\bibitem[\protect\citeauthoryear{{Reimers}}{{Reimers}}{2002}]{Reimers2002}
{Reimers} D.,  2002, \ssr, \href
  {https://ui.adsabs.harvard.edu/abs/2002SSRv..100...89R} {100, 89}

\bibitem[\protect\citeauthoryear{{Rost}, {Stasyszyn}, {Pereyra}  \&
  {Mart{\'\i}nez}}{{Rost} et~al.}{2020}]{Rost}
{Rost} A.,  {Stasyszyn} F.,  {Pereyra} L.,   {Mart{\'\i}nez} H.~J.,  2020,
  \mn@doi [\mnras] {10.1093/mnras/staa320}, \href
  {https://ui.adsabs.harvard.edu/abs/2020MNRAS.493.1936R} {493, 1936}

\bibitem[\protect\citeauthoryear{{Shandarin} \& {Klypin}}{{Shandarin} \&
  {Klypin}}{1984}]{Shandarin1984}
{Shandarin} S.~F.,  {Klypin} A.~A.,  1984, \sovast, \href
  {https://ui.adsabs.harvard.edu/abs/1984SvA....28..491S} {28, 491}

\bibitem[\protect\citeauthoryear{{Shandarin} \& {Zeldovich}}{{Shandarin} \&
  {Zeldovich}}{1989}]{Shandarin1989}
{Shandarin} S.~F.,  {Zeldovich} Y.~B.,  1989, \mn@doi [Reviews of Modern
  Physics] {10.1103/RevModPhys.61.185}, \href
  {https://ui.adsabs.harvard.edu/abs/1989RvMP...61..185S} {61, 185}

\bibitem[\protect\citeauthoryear{{Sousbie}}{{Sousbie}}{2011}]{disperse}
{Sousbie} T.,  2011, \mn@doi [\mnras] {10.1111/j.1365-2966.2011.18394.x}, \href
  {https://ui.adsabs.harvard.edu/abs/2011MNRAS.414..350S} {414, 350}

\bibitem[\protect\citeauthoryear{{Sousbie}, {Colombi}  \& {Pichon}}{{Sousbie}
  et~al.}{2009}]{Sousbie2009}
{Sousbie} T.,  {Colombi} S.,   {Pichon} C.,  2009, \mn@doi [\mnras]
  {10.1111/j.1365-2966.2008.14244.x}, \href
  {https://ui.adsabs.harvard.edu/abs/2009MNRAS.393..457S} {393, 457}

\bibitem[\protect\citeauthoryear{{Springel}}{{Springel}}{2005}]{Springel2005}
{Springel} V.,  2005, \mn@doi [\mnras] {10.1111/j.1365-2966.2005.09655.x},
  \href {https://ui.adsabs.harvard.edu/abs/2005MNRAS.364.1105S} {364, 1105}

\bibitem[\protect\citeauthoryear{{Tanimura}, {Aghanim}, {Bonjean}, {Malavasi}
  \& {Douspis}}{{Tanimura} et~al.}{2019}]{Tanimura2019}
{Tanimura} H.,  {Aghanim} N.,  {Bonjean} V.,  {Malavasi} N.,   {Douspis} M.,
  2019, arXiv e-prints, \href
  {https://ui.adsabs.harvard.edu/abs/2019arXiv191109706T} {p. arXiv:1911.09706}

\bibitem[\protect\citeauthoryear{{Tanimura}, {Aghanim}, {Bonjean}, {Malavasi}
  \& {Douspis}}{{Tanimura} et~al.}{2020}]{Tanimura2020}
{Tanimura} H.,  {Aghanim} N.,  {Bonjean} V.,  {Malavasi} N.,   {Douspis} M.,
  2020, \mn@doi [\aap] {10.1051/0004-6361/201937158}, \href
  {https://ui.adsabs.harvard.edu/abs/2020A&A...637A..41T} {637, A41}

\bibitem[\protect\citeauthoryear{{Tempel}, {Stoica}  \& {Saar}}{{Tempel}
  et~al.}{2013}]{Tempel2013}
{Tempel} E.,  {Stoica} R.~S.,   {Saar} E.,  2013, \mn@doi [\mnras]
  {10.1093/mnras/sts162}, \href
  {https://ui.adsabs.harvard.edu/abs/2013MNRAS.428.1827T} {428, 1827}

\bibitem[\protect\citeauthoryear{{Tempel}, {Stoica}, {Mart{\'{\i}}nez},
  {Liivam{\"a}gi}, {Castellan}  \& {Saar}}{{Tempel} et~al.}{2014}]{Tempel}
{Tempel} E.,  {Stoica} R.~S.,  {Mart{\'{\i}}nez} V.~J.,  {Liivam{\"a}gi} L.~J.,
   {Castellan} G.,   {Saar} E.,  2014, \mn@doi [\mnras]
  {10.1093/mnras/stt2454}, \href
  {http://adsabs.harvard.edu/abs/2014MNRAS.438.3465T} {438, 3465}

\bibitem[\protect\citeauthoryear{{Umehata} et~al.,}{{Umehata}
  et~al.}{2019}]{Umehata2019}
{Umehata} H.,  et~al., 2019, \mn@doi [Science] {10.1126/science.aaw5949}, \href
  {https://ui.adsabs.harvard.edu/abs/2019Sci...366...97U} {366, 97}

\bibitem[\protect\citeauthoryear{Walker et~al.,}{Walker
  et~al.}{2019}]{Walker2019}
Walker S.,  et~al., 2019, \mn@doi [Space Science Reviews]
  {10.1007/s11214-018-0572-8}, {215, 7}

\bibitem[\protect\citeauthoryear{{Welker} et~al.,}{{Welker}
  et~al.}{2020}]{Welker2020}
{Welker} C.,  et~al., 2020, \mn@doi [\mnras] {10.1093/mnras/stz2860}, \href
  {https://ui.adsabs.harvard.edu/abs/2020MNRAS.491.2864W} {491, 2864}

\bibitem[\protect\citeauthoryear{{Zel'dovich}}{{Zel'dovich}}{1970}]{Zeldovich1970}
{Zel'dovich} Y.~B.,  1970, \aap, \href
  {https://ui.adsabs.harvard.edu/abs/1970A&A.....5...84Z} {500, 13}

\makeatother
\end{thebibliography}



\bsp	
\label{lastpage}
\end{document}